\documentclass[usenatbib]{mnras}
\usepackage{textcomp}
\usepackage[utf8]{inputenc}
\usepackage{lscape}
\usepackage{lastpage}
\usepackage{amsmath,amssymb}
\usepackage{mathrsfs}	
\usepackage{multicol}
\usepackage[dvipdfmx]{graphicx}
\usepackage{caption}
\usepackage{subcaption}
\usepackage{multirow}
\graphicspath{{figures/}}

\usepackage{ulem}

\usepackage{times}
\usepackage{rotating}
\catcode`\@=11
\def\gta{\ifmmode{\,\mathrel{\mathpalette\@versim>\,}}
    \else{$\,\mathrel{\mathpalette\@versim>}\,$}\fi}
\def\lta{\ifmmode{\,\mathrel{\mathpalette\@versim<\,}}
    \else{$\,\mathrel{\mathpalette\@versim<}\,$}\fi}
\def\@versim#1#2{\lower 2.9truept \vbox{\baselineskip 0pt \lineskip
    0.5truept \ialign{$\m@th#1\hfil##\hfil$\crcr#2\crcr\sim\crcr}}}
\catcode`\@=12

{\newif\ifnotend
\notendtrue
\def\veclist{ABCDEFGHIJKLMNOPQRSTUVWXYZabcdefghijklmnopqrstuvwxyz.}
\def\top#1#2.{#1}
\def\tail#1#2.{#2.}
\loop\expandafter\xdef\csname bb\expandafter\top\veclist\endcsname%
{{\noexpand\bf\expandafter\top\veclist}}
\edef\veclist{\expandafter\tail\veclist}
\if\veclist.\notendfalse\fi\ifnotend\repeat}

\newcommand{\kpc}   {\,{\rm kpc}}
\newcommand{\pc}    {\,{\rm pc}}
\newcommand{\Mpc}   {\,{\rm Mpc}}
\newcommand{\Msun}  {\,M_{\odot}}
\newcommand{\Gyr}   {\,{\rm Gyr}}

\newcommand{\kms}   {\,{\rm km\,s^{-1}}}

\newcommand{\K}     {\,{\rm K}}
\newcommand{\h}     {\,{\rm h}}

\newcommand{\rhost}     {\rho_{\star}}

\newcommand{\rhodm} {\rho_{\rm dm}}
\newcommand{\Mdm}   {M_{\rm dm}}
\newcommand{\mdm}   {m_{\rm dm}}
\newcommand{\rhodmk}{\rho_{{\rm dm}, k}}

\newcommand{\Mvir}  {M_{\rm vir}}
\newcommand{\Mvirk} {M_{{\rm vir},k}}
\newcommand{\Tvir}  {T_{\rm vir}}
\newcommand{\rs}    {r_s}
\newcommand{\rhos}  {\rho_s}
\newcommand{\rvir}  {r_{\rm vir}}
\newcommand{\rhoc}  {\rho_{\rm c}}
\newcommand{\rc}    {r_c}
\newcommand{\MNFW}  {M_{\rm NFW}}
\newcommand{\McNFW} {M_{\rm coreNFW}}
\newcommand{\rhoNFW}{\rho_{\rm NFW}}

\newcommand{\Mst}   {M_{\star}}
\newcommand{\Mstk}  {M_{\star, k}}
\newcommand{\rhosii}{\rho_{-2}}
\newcommand{\rsii}  {r_{-2}}

\newcommand{\Ndm}   {N_{\rm dm}}
\newcommand{\Nst}   {N_{\star}}

\newcommand{\npart} {n_{\rm part}}

\newcommand{\dij}  {d_{ij}}
\newcommand{\xii}  {x_i}
\newcommand{\xj}   {x_j}
\newcommand{\xk}   {x_k}
\newcommand{\yi}   {y_i}
\newcommand{\yj}   {y_j}
\newcommand{\yk}   {y_k}
\newcommand{\zi}   {z_i}
\newcommand{\zj}   {z_j}
\newcommand{\zk}   {z_k}
\newcommand{\rk}   {r_k}

\newcommand{\Sij}  {S_{ij}}
\newcommand{\mk}   {m_k}
\newcommand{\rik}  {r_{i,k}}
\newcommand{\rjk}  {r_{j,k}}
\newcommand{\ak}   {a_k}

\newcommand{\Omegam}  {\Omega_{\rm m}}

\newcommand{\LL}    {\mathcal{L}}
\newcommand{\DD}    {\mathcal{D}}

\newcommand{\rhostk}{\rho_{\star, k}}

\newcommand{\lambdaiii} {\lambda_3}
\newcommand{\lambdaii}  {\lambda_2}
\newcommand{\lambdai}   {\lambda_1}

\newcommand{\rh}      {r_{\rm h}}

\newcommand{\Nbins} {N_{\rm bins}}

\newcommand{\dd}{\text{d}}

\newcommand{\Ho}    {H_0}

\newcommand{\qmean} {\langle q \rangle}
\newcommand{\smean} {\langle s \rangle}
\newcommand{\tmean} {\langle t \rangle}

\newcommand{\qbar}{\bar{q}}
\newcommand{\sbar}{\bar{s}}

\begin{document}

\date{}

\pubyear{2023}

\title[]{Shaping the unseen: the influence of baryons and environment on low-mass, high-redshift dark matter haloes in the SIEGE simulations}
{}
\author[R. Pascale et al.]{R. Pascale$^1$\thanks{E-mail: raffaele.pascale@inaf.it},
F. Calura$^1$, 
A. Lupi$^{2,3}$,
J. Rosdahl$^4$,
E. Lacchin$^{1,5}$,
M. Meneghetti$^1$,
C. Nipoti$^5$,
E. Vanzella$^1$,
\newauthor
E. Vesperini$^7$ and
A. Zanella$^6$
\\ \\
$^1$INAF - Osservatorio di Astrofisica e Scienza dello Spazio di Bologna, Via Gobetti 93/3, 40129 Bologna, Italy \\
$^2$DiSAT, Università degli Studi dell’Insubria, via Valleggio 11, I-22100 Como, Italy \\
$^3$INFN, Sezione di Milano-Bicocca, Piazza della Scienza 3, 20126 Milano, Italy \\
$^4$Centre de Recherche Astrophysique de Lyon UMR5574, Univ Lyon, Univ Lyon1, Ens de Lyon, CNRS, F-69230 Saint-Genis-Laval, France \\
$^5$Dipartimento di Fisica e Astronomia “Augusto Righi” – DIFA, Alma Mater Studiorum – Università di Bologna, via Gobetti 93/2, I-40129 Bologna, Italy \\
$^6$Istituto Nazionale di Astrofisica, Vicolo dell’Osservatorio 5, I-35122 Padova, Italy \\
$^7$Department of Astronomy, Indiana University, Bloomington, Swain West, 727 E. 3rd Street, IN 47405, USA}
\label{firstpage}
\pagerange{\pageref{firstpage}--\pageref{lastpage}}

\maketitle
 
\begin{abstract}
We use zoom-in, hydrodynamical, cosmological $N$-body simulations tracing the formation of the first stellar clumps from the SImulating the Environments where Globular clusters Emerged (SIEGE) project, to study key structural properties of dark matter haloes when the Universe was only $0.92\Gyr$ old. The very high-resolution (maximum physical resolution $0.3\h^{-1}\pc$ at $z=6.14$, smallest dark-matter particle mass $164\Msun$) allows us to reach the very low mass end of the stellar-to-halo mass relation ($\Mvir=10^{7.5-9.5}\Msun$) to study the processes that mould dark matter haloes during the first stages of structure formation. We investigate the role of baryonic cooling and stellar feedback, modeled from individual stars, in shaping haloes, and of environmental effects as accretion of dark matter along cosmic filaments and mergers. We find that the onset of star formation (typically for $\log\Mvir/\Msun\simeq7.6$) causes the inner cusp in the haloes’ density profile to flatten into a core with constant density and size proportionally to the halo virial mass. Even at these mass scales, we confirm that baryons make haloes that have formed stars rounder in the central regions than haloes that have not formed stars yet, with median minor-to-major $\qmean$ and intermediate-to-major $\smean$ axes 0.66 and 0.84, respectively. Our morphological analysis shows that, at $z=6.14$, haloes are largely prolate in the outer parts, with the major axis aligned along filaments of the cosmic web or towards smaller sub-haloes, with the degree of elongation having no significant dependence on the halo mass.
\end{abstract}

\begin{keywords}
galaxies: haloes - galaxies: high-redshift - cosmology: early Universe - galaxies: formation - galaxies: kinematics and dynamics - galaxies: structure
\end{keywords}

\maketitle

\section{Introduction}
\label{sec:intro}

According to the current cosmological model, dark-matter represents a crucial and significant component of almost all complex systems that populate the Universe. It fills and shapes the cosmic web, it wraps the tiniest dwarf galaxies up to the largest galaxy clusters, it governs the dynamics and behavior of cosmic structures on nearly all scales \citep{PlankXIII}. All of these systems stem from the evolution of high-density fluctuations in the primordial density field, and exhibit a non-linear, hierarchical growth through the merging of smaller structures and the accretion of matter along filaments \citep{Lacey1993,Lacey1994}. In this scenario, it is natural to expect that the structural and kinematic characteristics of dark matter haloes may depend on the environment where they live in \citep{Allgood2006,VeraCiro2011}, on the particles they are made of \citep{Spergel2000,Hui2017,Nadler2021}, on the baryonic matter they (may) host \citep{Bullock2005} and their overall mass assembly history. 

While, for instance, the central parts of haloes, where galaxies originate, may be more sensitive to processes involving infall of baryons, baryonic cooling and stellar feedback \citep{Kazantzidis2004,Abadi2010,Butsky2016}, the structure of the outer haloes is, instead, expected to be driven by interactions with systems of comparable or lower mass or by ramifications of the cosmic web \citep{Maccio2008,Tomassetti2016}. In this respect, the three-dimensional shape and orientation of haloes, and the presence of cores with constant densities at their centers can provide crucial information on these properties and aid in understanding the evolution and assembly history of galaxies \citep{Dubinski1991,Cole1996}.


In the $\Lambda$CDM model, dark matter haloes assemble hierarchically \citep{White1978}, with massive systems forming through the merging of smaller ones. These mergers are inherently clumpy, directional, and anisotropic. As a result, haloes should not be perfectly spherical, which is, for instance, an assumption made in the analytical top-hat spherical collapse model \citep{Gunn1972, Gunn1977} commonly used to describe halo formation. This is particularly important considering that the relaxation time of haloes is often longer than the typical timescale required for mergers to occur \citep{Allgood2006}. The vast majority of predictions of halo shapes comes from $N$-body simulations, both in the form of dark matter-only simulations \citep[hereafter DMO; ][]{Dubinski1991,Allgood2006,Maccio2008,JeesonDaniel2011}, where the baryonic component is neglected, or hydrodynamical ones \citep{Bryan2013,Prada2019,Chua2019,Chua2022}. All these research works show, indeed, that haloes deviate significantly from spherical symmetry, to a degree that depends on the halo mass, on the redshift, on the presence or absence of stars and gas and on the stellar feedback model implemented in the simulation. While early studies of halo morphology produced conflicting results \citep{Frenk1988,Dubinski1991,Warren1992}, nowadays it is reasonably well established that present-day halo shapes vary {at least} with the distance from the halo center, with a tendency to be triaxial/oblate in the outer parts \citep{Abadi2010,VeraCiro2011,Chua2019}. Moreover, several authors found that the inclusion of baryons makes haloes distinctly more spherical, especially in the halo center. For instance, by studying Milky Way (MW)-sized galaxies ($\Mvir\simeq10^{12}\Msun$) shapes in the Illustris simulations \citep{Genel2014,Vogelsberger2014b,Vogelsberger2014}, and comparing them with similar haloes in DMO simulations, \cite{Chua2019} found that the average minor-to-major axis of these haloes increases from $\qmean=0.52\pm0.10$ in DMO to $\qmean=0.70\pm0.11$ in full physics simulations, and the intermediate-to-major axis $\smean$ from $0.67\pm0.14$ to $0.88\pm0.10$ (see also \citealt{Bryan2013,Butsky2016,Prada2019,Cataldi2021} for similar results, but based on different suites of simulations). This 'sphericization' \citep{Dubinski1994} of the central haloes induced by baryons somewhat reduces the tension between previous results from simulations and indirect measures of MW-like halo shapes obtained from orbital analysis of stellar streams, such as the Sagittarius stream \citep{Ibata2001,VeraCiro2013}, Pal 5 and GD--1 \citep{Bovy2016}, or from the analysis of high velocity stars \citep{Law2010}.



However, while theoretical models and observations at $z=0$ confirm the presence of non-spherical shapes in dark matter haloes, our current understanding of the process by which these shapes are acquired remains incomplete. This is especially true when considering two crucial factors: i) the influence of the dynamics of the surrounding large-scale structure in the non-linear regime, and ii) the incorporation of the formation and co-evolution of baryonic structures within dark haloes. With only a few exceptions, these factors have been insufficiently investigated, as, for instance, the majority of previous studies on halo shapes have predominantly focused on the present-day structural properties of massive haloes ($\Mvir\gtrsim10^{10}\Msun$). Following the shape evolution with redshift of haloes from DMO simulations in the standard cosmological model, \cite{Allgood2006} found that, at redshift $z=0$, the oldest haloes tend to become spherical earlier, and more rapidly than haloes forming later in time, almost independently of the final mass. Focusing on five MW-like galaxies at $z=0$ in the Aquarius simulations, \cite{VeraCiro2011} also recognizes environment effects as the primary driver of the redshift evolution of halo shapes, with haloes being more prolate at high redshift than at $z=0$. Therefore, analyzing the shapes of haloes, from high to low redshifts, and reconstructing the halo assembly history can provide insights into the nature of dark matter and the processes involved in halo and galaxy formation.

As mentioned, the presence of cusps or cores of constant densities at the centres of dark matter haloes can also have a significant impact in differentiating between cosmological and galaxy evolution models \citep{Bullock2017,DelPopolo2022}. The contrast between DMO simulations, that predict haloes to accumulate mass in the inner parts following cuspy profiles, and measurements based on observations of rotation curves of local disc galaxies is, indeed, an indication that the effects of baryons may not be negligible on small scale halo structures \citep{Pontzen2012} or may alternatively hint that exotic dark matter models / cosmologies may be required \citep{Governato2015}. As for baryon-induced cores, it has been shown that they can be formed in a cosmological context in various ways: before star formation in the halo ignites, by heating the cusp via dynamical friction of fragmented gas \citep{NipotiBinney2015}; by subsequent and rapid events of star formation if enough energy is transferred to dark matter \citep{NavarroEF1996,Governato2012,Read2005}. Although these processes are persuading, much still remains debated as, for instance in the latter case, the halo/stellar threshold mass for an efficient core formation \citep{Governato2012,DiCintio2014}, or the dependence of the simulations on the resolution and on the stellar feedback \citep{Vogelsberger2014c}.

The results emerging from these investigations indicate that mechanisms related to density sphericization and core-cusp transformation manifest primarily in the central regions of haloes, where the fraction of baryonic-to-dynamical mass is the highest. Moving towards the outer parts of haloes, where this fraction becomes much smaller, the environment becomes increasingly more important. This radial distribution of baryonic-to-dark mass, however, varies as a function of halo mass and redshift, and the occurrence and impact of these processes on low-mass, high-redshift haloes remain poorly understood. In this work, we rely on state-of-the-art, high-resolution cosmological simulations to study the properties of young dark matter haloes during the first stages of cosmic structure formation. We analyze the least massive haloes to form, pushing towards the very low mass end of the high redshift stellar-to-halo mass relation. Our very high spatial resolution and the ability to resolve stellar feedback from individual stars allow us to inspect the complex interplay between dark matter haloes and baryons, how and to what extent the former influence the latter in a Universe that is just $0.92\Gyr$ old, and the role of environment in shaping the structures that will end up dominating present-day galaxies and clusters. To our knowledge, this represents the first time that the effects of baryons and environment on the haloes' inner and outer structures are studied on these low mass scales and high redshift.

The paper is organized as follows: in Section~\ref{sec:simu}, we describe the set-up of the simulations analyzed in this work; in Sections~\ref{sec:id} and~\ref{sec:struct} we describe the methods used to identify dark matter haloes and stellar clumps in the simulations, and the fitting procedures adopted to derive relevant quantities such as halo centers, density profiles, and shapes. In Section~\ref{sec:res}, we discuss our results and compare them with the literature. Finally, in Section~\ref{sec:conc} we draw our conclusions.

\section{Simulation}
\label{sec:simu}
The first simulation analyzed in this work is a part of the SImulating the Environment where Globular clusters Emerged (SIEGE) project, a suite of hydrodynamical, cosmological, zoom-in simulations aimed at a detailed study of the formation of the first star-forming clumps in the Universe, that includes sub-parsec resolution and feedback from individual stars.  This specific simulation, first presented in \cite{Calura2022}, is aimed at modelling the properties of an extended star-forming complex observed in the lensed field of the galaxy cluster MACS J0416.1–2403, at redshift $z=6.14$ \citep{Vanzella2019,Calura2021}. The simulation was performed with the adaptive mesh refinement (AMR) hydrodynamic code \textsc{ramses} \citep{Teyssier2002} and evolved down to $z=6.14$, when the Universe was $0.92\Gyr$ old. In this work, we adopt the same $\Lambda$CDM, flat cosmological model as in the simulation, with matter density $\Omegam = 0.276$ and Hubble constant $\Ho = 70.3\kms\Mpc^{-1}$ \citep{Sharov2014,Omori2019}. The maximum resolution allowed by the use of AMR is $0.3\h^{-1}\pc$ at $z=6.14$. 

The physical processes included in the simulation are: i) atomic radiative cooling due to hydrogen, helium and metals in photoionisation equilibrium with a redshift-dependent ionizing ultraviolet (UV) background \citep{Haardt2012}; ii) formation of individual stars; iii) consequent stellar winds (SWs) and supernovae (SNe) feedback modeled directly for individual stars. To avoid artificial radiative loss of the energy injected by the stars, a delayed cooling feedback scheme as described in \cite{Teyssier2013} is adopted in the simulation. 
\begin{figure*}
    \centering
    \includegraphics[width=0.475\hsize]{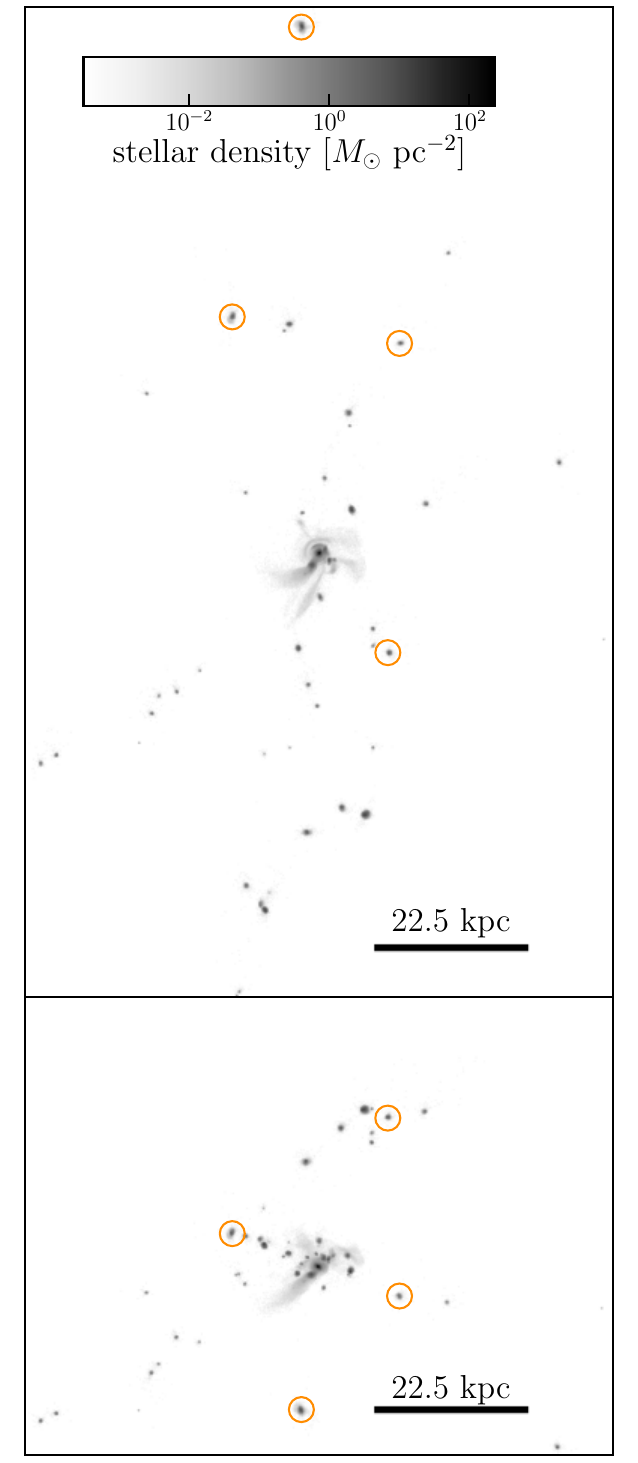}
    \includegraphics[width=0.475\hsize]{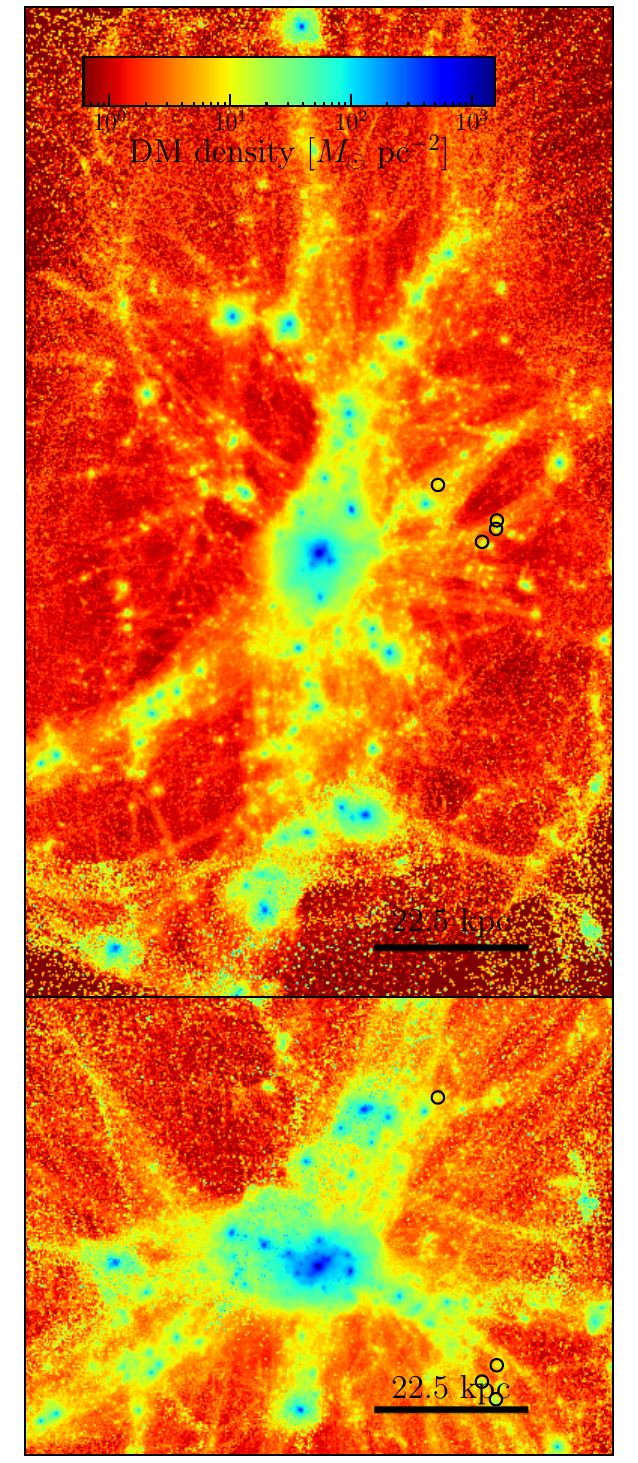}
    \caption{Left-hand panels: stellar surface density projected along the $y$ direction (top) and the $z$ direction (bottom). Right-hand panels: same as the left panels, but showing the dark-matter surface density map. In the right panels, blue colors correspond to high-density regions while red colors to low-density regions. The orange circles in the left panels show the positions of clumps shown in the top and middle panels of Fig.~\ref{fig:stdmprof}, while the black circles in the right panels mark the position of starless haloes later analyzed in the bottom panels of Fig.~\ref{fig:stdmprof}.}
    \label{fig:stdm}
\end{figure*}
In particular, stars in the mass range $[3,8]\Msun$ give a dominant contribution in terms of ejected mass in the AGB phase, although they are not a substantial source of heating for the gas; stars in the mass range $[8,40]\Msun$ (massive stars) contribute with both pre-SN and SN feedback while more massive stars collapse instantaneously, not polluting the system. In all cases, the amount of mass, energy and metals deposited in the ISM depends on the stars initial mass. Stars are created individually by means of a IMF sampling method \citep{Sormani2017,Andersson2020}: for the adopted IMF \citep[][]{Kroupa2001}, the mass range is divided in 12 bins and, when a gas cell reaches the condition for star formation and a stellar particle is formed, individual stars are drawn from the IMF via Poisson sampling. To avoid the computational expense of too many stellar particles, all individual stars with mass $<3\Msun$ are grouped in single star particles. Since the simulation is evolved for less than $1\Gyr$, these star particles do not contribute in terms of stellar feedback as they live longer than the age of the Universe at $z=6.14$. The choice to generate individual stars is motivated by the very high resolution reached by the simulation that, in most cases, makes the amount of gas eligible for star formation enough to generate a few stars only. 

The initial conditions (ICs) of the simulation have been generated via the \textsc{music} software \citep{Hahn2011} at $z = 100$. To define the zoom-in region, the simulation box is initially described by $64^3$ dark matter particles with $64^3$ cells and the simulation is run down to $z=6.14$. Then, an isolated halo with a mass within three times its virial radius of $\simeq4\times10^{10}\Msun$ is identified with the \textsc{hop} halo finder \citep{Eisenstein1998}. All the particles belonging to the target halo are flagged, traced back in time, and two refinement levels are added. The whole procedure is repeated another time and the ICs are computed again with the inclusion of baryons. In the end, the initial number of dark matter particles in the simulation is $2\times10^8$, with a lowest dark matter mass per particle of $164\Msun$. The procedure adopted is the same as in \cite{Fiacconi2017} and \cite{Lupi2019}. For further details on the generation of the ICs, the implementation of star formation, feedback or cooling processes we refer to \cite{Calura2022}.

To check whether our results are driven (or not) by the inclusion of a baryonic component in the simulation, we include in all the subsequent analysis also a second simulation, which we will refer to as DMO, that comprises solely dark matter, that is set up with the same ICs, and that is evolved down to the same redshift as the full physics simulation just described.

\section{haloes and stellar clumps identification}
\label{sec:id}
We now outline the algorithm adopted to locate dark matter haloes and stellar clumps (Sections~\ref{subsec:dmids} and \ref{subsec:stids}), and the methods used to build the samples of haloes and stellar systems in the full physics simulation (Section~\ref{subsec:crossmatch}) and the sample of haloes in the DMO simulation (Section~\ref{subsec:dmo}).

\subsection{Dark matter}
\label{subsec:dmids}
To identify dark matter haloes in the full physics simulation, we rely on the density-based, hierarchical clustering method proposed by \cite{Campello2013}, as implemented in the software library \textsc{hdbscan} \citep{McInnes2017}\footnote{\url{https://hdbscan.readthedocs.io/en/latest/index.html}}. \textsc{hdbscan} finds overdensities and groups in multi-dimensional sets of data, minimizing a predetermined distance between elements of the group. One of the advantages of \textsc{hdbscan} with respect to the classical \textsc{dbscan} \citep{Ester1996,Schubert2017} is that it does not require to specify a maximum distance ($\epsilon$) used as boundary to define if two elements of the same dataset are part of the same group, but it rather marginalizes over $\epsilon$, evaluating the best value for each cluster. The algorithm thus adapts to the local density of multi-dimensional data and it performs much better on datasets where clusters are expected with very different densities, as in our specific case. 

We select all dark matter particles within a cuboid enclosing the central clump and the filament that branches out of it, and we run \textsc{hdbscan} 
adopting as metric the Euclidean distance
\begin{equation}\label{for:Euc}
    \dij \equiv \sqrt{(\xii-\xj)^2 + (\yi-\yj)^2 + (\zi-\zj)^2},
\end{equation}
where the triplet $\{\xk, \yk, \zk\}$ denotes the coordinates of the $k$-th dark matter particle. As a reference, the right panels of Fig.~\ref{fig:stdm} show the dark matter surface density within the selection box when the system is projected along two different directions. As relevant parameter, \textsc{hdbscan} only demands to specify the minimum number of elements per group ($\npart$) which we set to $\npart=50000$. Since all dark-matter particles within the region of interest have the same mass ($\mdm=164\Msun$), this requirement approximately fixes the minimum halo mass that the algorithm will sample to $\simeq8.2\times10^6\Msun$. The total number of dark-matter haloes thus identified is 190.

\subsection{Stars}
\label{subsec:stids}
We repeat the same procedure running \textsc{hdbscan} over the stellar particles, but excluding those particles lying within a cube $\simeq11\kpc^3$ wide, centered on the central clump, and we set $\npart=250$. Differently from halo particles, stellar particles have different mass. However, since different stellar particles are merged into one if, individually, less massive than $3\Msun$, the minimum stellar mass sampled by the algorithm is $750\Msun$. Nevertheless, the algorithm identifies the same number of clumps almost independently from $\npart$: as it is particularly evident from the top-left panel of Fig.~\ref{fig:stdm}, where we show the projected stellar mass density of the same portion and along the same line-of-sight as in the top-right panel, the separation between stellar clumps in the full three dimensional space is very clear.


The central region excluded from the clustering algorithm is rich of substructures: i) the massive and dense roundish clump at the center of the box; ii) the stellar stream that wraps the central clump to the north; iii) at least four smaller systems that are merging with the central clump. When included in the clustering algorithm, this portion of the simulation box is unphysically fragmented into a relatively large number of clumps whose number depends on $\npart$ and that, from a simple investigation by eye, overlap or are poorly distinct between one another. This is why we exclude this portion from the automatic stellar clump identification via \textsc{hdbscan} and manually add the four aforementioned clumps to the ones identified via \textsc{hdbscan}. The total number of stellar clumps identified in this way is 45. We note that while we manually add these four clumps to the sample, the dark matter haloes they live in have all been identified by the clustering algorithm run over the dark-matter particles.

\subsection{Building the final samples}
\label{subsec:crossmatch}
We determine the centers of the stellar clumps via the shrinking sphere method \citep[SSM; ][]{Power2003}, an iterative scheme used to evaluate the center of mass of a given set of particles with known masses and positions. At the iteration $i+1$, the algorithm computes the center of mass considering all the particles enclosed within a sphere centered on the $i$-th estimate of the center of mass, and with radius $r_{i+1}=K r_i$, with $r_i$ the radius of the sphere at the $i$-th iteration and $0<K<1$ a constant. The algorithm stops when the sphere contains a predetermined number of particles. In all our computations we set $K=0.95$. 

We consider all particles classified as group members by \textsc{hdbscan} so that, at the zeroth iteration, the initial sphere has a diameter equal to the largest distance among any pair of star particles in the group. The algorithm stops when the sphere is left with $\simeq$5\% of those particles. In the case of the four additional clumps that have not been identified by \textsc{hdbscan}, at the zeroth iteration, we consider all particles enclosed within a sphere whose center lies within the clump and with $\simeq300\pc$ radius. We tested the algorithm against different stop conditions, specifically 2.5\% and 1\% of the initial particles, finding converged results, i.e. with a relative variation of the center of mass with respect to the 5\% case always smaller than $10\pc$ for the majority of the clumps, and smaller than $40\pc$ in a few cases. 


For these 45 clumps only, we run SSM over the dark matter particles to get a first estimate of the center of the haloes in which they are embedded. We use the center of the stellar component previously determined as initial guess, and we consider all dark-matter particles within the stellar clump half-mass radius, with the stop condition set to 2\% of the dark-matter particles within the initial sphere. We again tested the method against stop conditions of 1\% and 5\% of the input dark-matter particles finding difference between the centers, in most cases, always smaller than $15\pc$.

For the dark matter haloes identified with \textsc{hdbscan}, we use the same iterative scheme and consider the dark matter particles identified as group members by the clustering algorithm. Here, SSM stops when 1000 particles are left inside the sphere. We later repeat the process starting from the previous estimate of the center of mass and passing all dark-matter particles within a $1\kpc$ sphere radius, not just the ones flagged as members.

Once the centers of the dark matter haloes, identified both by the clustering algorithm and as the dark counterparts of the stellar clumps, are determined, we cross-match the samples to eliminate any duplicates. We consider as duplicates all haloes whose centers of mass measured by the two methods are closer than $200\pc$ and, in these cases, we take as center of mass the one computed from the stars. The total number of haloes after the cross-match is 195\footnote{Five dark-matter haloes with a stellar counterpart were not detected by the clustering algorithm when finding groups of dark matter particles.}.

\subsection{The DMO simulation}
\label{subsec:dmo}

We run the same exact procedure of halo identification described in Section~\ref{subsec:dmids} on the simulation comprising only dark matter (same $\npart$ and same spatial cut of the simulation box) finding a total of 194 dark-matter haloes. In this case, we do not need to make any cross match and we just determine the halo centers with the SSM method in the same way as we did with the haloes without a stellar counterpart in the simulation with baryons.


\section{Structural properties of dark haloes and stellar clumps}
\label{sec:struct}

Here, we explain the methods used to determine important structural properties such as shape, stellar mass, size, virial mass, and density distributions of dark matter haloes and stellar clumps. For clarity, we describe the general algorithm used to calculate the shape of any three-dimensional distribution of particles in Section~\ref{subsec:shape}, and we separately analyze the dark matter haloes of both simulations in Section~\ref{subsec:dm} and the stellar components in Section~\ref{subsec:stars}.

\subsection{Shape computation}
\label{subsec:shape}
To compute the shape of any given target (i.e. the directions and elongations of the principal axes of a set of dark matter or stellar particles) we diagonalize the system's unweighted shape tensor \citep[e.g.][]{Zemp2011}
\begin{equation}\label{for:shape}
    \Sij \equiv \frac{1}{\sum_{k=1}^N \mk}\sum_{k=1}^N \mk\rik\rjk,
\end{equation}
where $\mk$ is the mass of the $k$-th particle, $\rik$ and $\rjk$ are the $i$-th and $j$-th components of its position vector and $\Sij$ is the $ij$ element of the shape tensor. The sum in equation (\ref{for:shape}) extends over all particles within a triaxial shell of major axis width $\Delta a\equiv a_{l+1}-a_l$, where
\begin{equation}\label{for:rlim}
    a_l \le \sqrt{\xi^2 + \frac{\eta^2}{s^2} + \frac{\zeta^2}{q^2}} \le a_{l+1},
\end{equation}
with $(\xi,\eta,\zeta)$ the coordinates in the principal frame, and 
\begin{equation}
    s\equiv b/a \quad \text{and} \quad q=c/a
\end{equation}
the triaxial shell intermediate-to-major ($b$ to $a$) and minor-to-major ($c$ to $a$) axes. While the eigenvectors of the shape tensor represent the shell's principal axes (i.e. they are proportional to $\xi$, $\eta$ and $\zeta$), called $\lambdai>\lambdaii>\lambdaiii$ the eigenvalues of the shape tensor, then
\begin{equation}
    s\equiv\sqrt{\frac{\lambdaii}{\lambdai}},\quad q\equiv\sqrt{\frac{\lambdaiii}{\lambdai}}.
\end{equation}
In our analysis, the eigenvalues and eigenvectors are computed iteratively starting from the guess $s=q=1$, which corresponds to purely spherical bins. The algorithm stops when the relative variation of $q$ and $s$ between subsequent iterations is less than 1\%.

In order to guarantee that even the least massive clumps are sampled with a sufficiently large number of particles, for the stellar structures we do not consider variations of shape with the distance from the center and, for the shape computation, all stellar particles are grouped within one shell only. On the contrary, in the case of the dark haloes (both in the DMO and full physics simulations), we bin the particles in ellipsoidal shells of semi-major axes evenly spaced in log-scale. The bins extend, approximately, from tens of $\pc$ to $4-5\kpc$ and are, in number, proportional to $\ln^{\frac{6}{5}}\Ndm$, a function that, empirically, we find to ensure a good sampling of the density profiles, where $\Ndm$ is the number of dark-matter particles within $1\kpc$. Only for the haloes that host stars, we further merge in a single bin all those dark-matter particles within the radial extent of the corresponding stellar component. 


\subsection{Dark matter density profiles}
\label{subsec:dm}

For any given halo, we compute the dark-matter density distribution in the same bins used to determine the shape profiles. In each radial bin, we rotate the dark-matter particles into a reference frame whose axes are aligned with the bin's principal axes. The particles are then split into 12 cloves of constant volume and the density profile of the halo is computed in each clove. We take the average density of the 12 cloves as a measure of the density profile and its dispersion as the error. With this procedure, the errorbars associated to each profile measure deviations from triaxiality. Note that, in the case of haloes hosting stars, the inner density distribution is computed using a fixed shape (see Section~\ref{subsec:shape}). 

We fit the binned density profile of each halo with the coreNFW model \citep{Read2016}
\begin{equation}\label{for:mcnfw}
    \McNFW(< r) = \MNFW(< r)f^n,
\end{equation}
with
\begin{equation}\label{for:rc}
    f\equiv \tanh\biggl(\frac{r}{\rc}\biggr).
\end{equation}
In equation (\ref{for:mcnfw}), $\McNFW$ is the mass profile of the coreNFW, while $\MNFW$ is the mass enclosed within the classical \citet[hereafter NFW]{NavarroFrenkWhite1996} model
\begin{equation}\label{for:MNFW}
    \MNFW(<r) = 
     4 \pi \delta \rhoc \rs^3 \biggl[\ln\biggl(1 + \frac{r}{\rs}\biggr) - \frac{r}{\rs} \biggl(1 + \frac{r}{\rs}\biggr)^{-1}\biggr],
\end{equation}
which corresponds to the density distribution
\begin{equation}\label{for:rhoNFW}
    \rhoNFW(r) =\frac{\delta\rhoc}{\frac{r}{\rs}\biggl(1+\frac{r}{\rs}\biggr)^2}.
\end{equation}
In equations (\ref{for:MNFW}) and (\ref{for:rhoNFW}), $\rs$ is the halo scale length while $\rhoc$ is the critical density of the Universe at the time of the simulation. Also,
\begin{equation}
    \delta \equiv \frac{\Delta}{3}\frac{c^3}{\ln(1+c)-\frac{c}{1+c}},
\end{equation}
with $c$ the halo concentration. Dark matter haloes are usually described in terms of virial radius $\rvir$ and virial mass $\Mvir$. The virial radius is commonly assumed as the distance where the halo average density is $\Delta(=200)$ times $\rhoc$, while the virial mass is $\Mvir\equiv\MNFW(<\rvir)$. The halo concentration is, then, $c\equiv\rvir/\rs$. At $z=6.14$, in a flat $\Lambda$CDM Universe with the adopted $\Omegam=0.276$ and $\Ho=70.3\kms\Mpc^{-1}$, $\rhoc=1.389\times10^{-5}\Msun\pc^{-3}$. The coreNFW behaves as an NFW at large radii, but it allows to describe possibly cored density distributions on scales $r<\rc$ (equation \ref{for:rc}). The strength of the core is controlled by the dimensionless parameter $0\le n\le1$: when $n=1$, the core has its maximum strength and the transition between core and outer regions is sharp, whereas when $n=0$ the model is an NFW. 


\begin{figure*}
    \centering
    \includegraphics[width=.475\hsize]{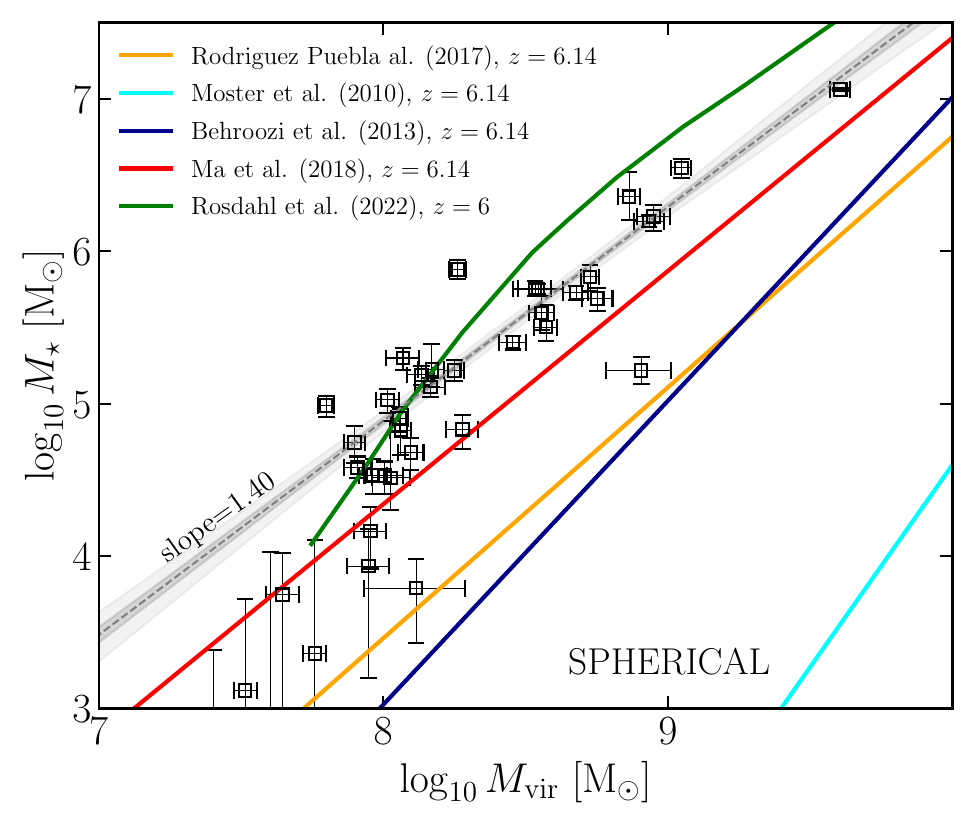}
    \includegraphics[width=.475\hsize]{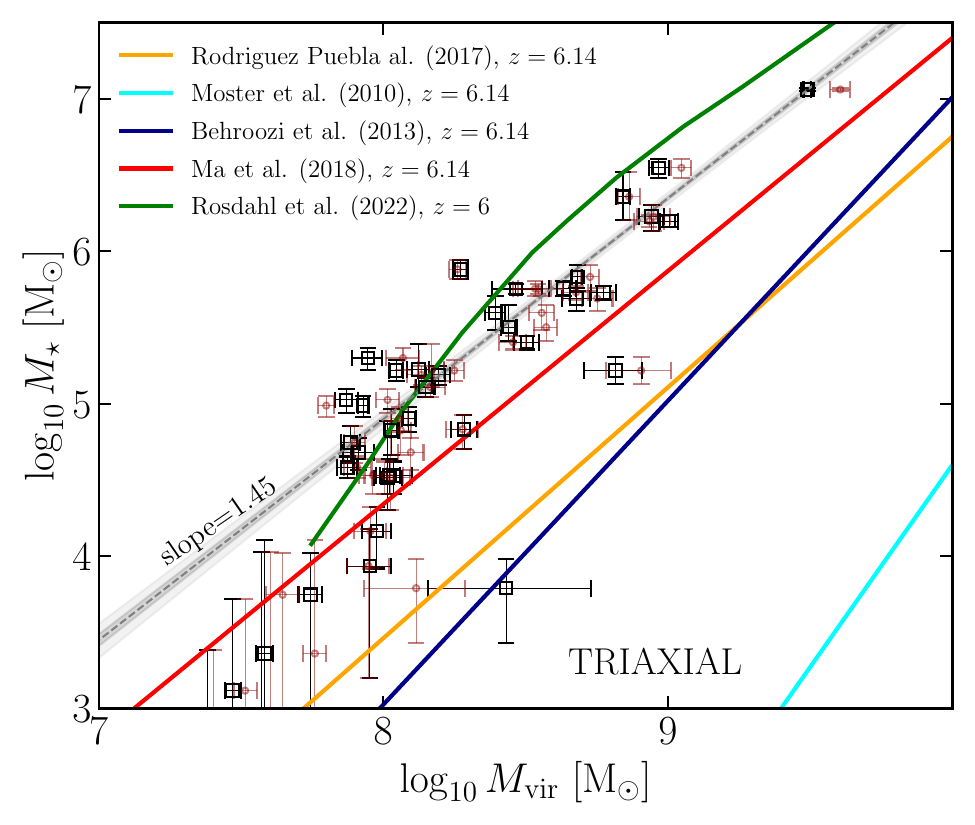}
    \caption{Left panel: stellar-to-halo-mass relation obtained computing the halo virial masses using spherical bins (black squares with errorbars). Points are shown alongside with the median (dashed grey line), $1\sigma$ and $3\sigma$ models (grey bands) derived via a linear fit of median slope $1.40\pm0.3$ and $q=-6.35\pm0.27$. For a comparison, the red solid line is the SHMR from \citet{Ma2019}; the orange line from \citet{RodriguezPuebla2017}; the blue line from \citet{Behroozi2013}, the cyan from \citet{Moster2010} and the green line from \citet{Rosdahl2022}. Right panel: same as the left panel but here the halo virial masses are computed accounting for the triaxial shape of the haloes. In this case the median slope of the relation is $1.45\pm0.02$ while $q=-6.70_{-0.18}^{+0.17}$. Here, we added the spherically binned haloes as light red points.}\label{fig:SHMR}
\end{figure*}

We run a Markov Chain Monte Carlo (MCMC) method to explore the parameter space and to sample the model's posterior distribution used to define the best fit parameters and confidence intervals. We run 16 chains per halo adopting uniform priors on the logarithms of $\rhos\equiv\rs^3\delta\rhoc$, $\rs$, $\rc$ and uniform on $0<n<1$, with $\bxi\equiv\{\rhos,\rs,\rc,n\}$ the model's free parameter vector. To sample from the posterior, we use a combination of the differential evolution proposal by \cite{Nelson2014} and the snooker proposal by \cite{terBraak2008} as implemented in the software library \textsc{emcee} \citep{ForemanMackey2013}. The log-likelihood of the model given the data $\DD$ is
\begin{equation}\label{for:logl}
    \ln\LL(\bxi|\DD) = -\frac{1}{2}\sum_{k=1}^{\Nbins}\biggl[\frac{\rhodm(\ak)-\rhodmk}{\delta\rhodmk}\biggr]^2,
\end{equation}
where $\DD\equiv\{\ak,\rhodmk,\delta\rhodmk\}$ is the halo binned density profile previously derived, $\rhodm$ the density of the coreNFW and $\Nbins$ is the number of radial bins. In the fitting procedure we exclude from the profiles all radial bins less dense than $40\rhoc$. Depending on the specific halo, we always remove a burn-in of at least 1500 steps and adopt a thinning of $<40$ steps, always of the order of the chains' auto-correlation length. The remaining steps are used to build the posterior distributions over $\bxi$. All the uncertainties over the models parameters or any derived quantity are estimated as the 16th, 50th and 84th percentiles of the corresponding distributions.

Since we consider that the dark matter mass distribution ($\Mdm$) stratifies on triaxial shells of semi-major axis $a$, but equations from (\ref{for:mcnfw}) to (\ref{for:rhoNFW}) are valid for spherically symmetric models, the actual dark matter mass enclosed within $a$ is related to that inferred from the model by
\begin{equation}\label{for:realmass}
    \Mdm(<a) \equiv \qbar\sbar\McNFW(<a).
\end{equation}
In the above equation $\qbar\sbar$ is the product of the minor-to-major and intermediate-to-major axes of the halo and it must be assumed constant. We determine the halo virial radius $\rvir$ from the fitting procedure\footnote{In our case $\rvir$ is the semi-major axis where the average triaxial density is $200\rhoc$.} and we later use that estimate to compute the average shapes $\qbar$ and $\sbar$ directly from the simulation, considering all dark matter particles within $0.75\rvir$. We then define the virial mass of our haloes as
\begin{equation}
   \Mvir = \qbar\sbar\McNFW(\rvir).
\end{equation}
As we will discuss in Section \ref{subsec:barshapes}, halo shapes do not depend significantly on the distance from the center, which justifies the use in equation~(\ref{for:realmass}) of fixed shape for the virial mass computation.

At the end of the entire procedure of shape determination and density computation, an additional 39 haloes were removed from the sample since they resulted in unphysical halo parameters (e.g. $c<1$, $q\simeq s\simeq0$) due to the misclassification of portions of the background or tight interacting haloes. 6 of these haloes where hosting stars. This lowers the total number of haloes used to 156 and, consequently, the total number of stellar clumps included in the analysis to 39.

The entire fitting procedure is repeated for the haloes identified within the DMO simulation. In this case we removed 35 haloes lowering the total number of haloes analyzed to 159.

\subsection{Stellar density profiles}
\label{subsec:stars}

The stellar density profiles are also computed in radial bins of triaxial shape, with all bins having the same shape. The bins extend out to a few half-mass radii, estimated directly from the stellar particles distribution, and they are spaced such that the distance between the arc-tangent of two adjacent edges is constant, and are, in number, proportional to $\ln^{\frac{6}{5}}\Nst$, with $\Nst$ the total number of stellar particles within the previous estimate of the half-mass radius. 

We fit the clump density profiles with the \cite{Einasto1965} model
\begin{equation}\label{for:einasto}
    \rhost(r) = \rhosii \exp{\biggl\{-2n\biggl[\biggl(\frac{r}{\rsii}\biggr)^{\frac{1}{n}}-1\biggr]\biggr\}},
\end{equation}
where $\rsii$ is the radius where the logarithmic slope $\frac{\dd\ln\rhost}{\dd \log r}|_{\rsii}=-2$, $\rhosii$ the corresponding density. Here $n$ controls the shape of the inner and outer profile. We run the same MCMC procedure used for the dark matter haloes, but using 12 chains per clump. The log-likelihood of the model is the same as in equation (\ref{for:logl}), but $\bxi\equiv(\rhosii,\rsii,n)$, $\DD=\{\rk,\rhostk,\delta\rhostk\}$ is the stellar binned density profile and the model is given by equation (\ref{for:einasto}). We use uniform and wide priors on the logarithms of $\rhosii$ and $\rsii$, while we restrict to $0.2<n<10$. Also in this case, since stars stratify on triaxial shells, but equation~(\ref{for:einasto}) is that of a spherical model, the total stellar mass is
\begin{equation}
    \Mst=4\pi q s \int_0^{+\infty}\rhost(a)a^2\dd a,
\end{equation}
with $q$ and $s$ the (constant) semi-minor and semi-major axis lengths determined in Section~\ref{subsec:shape}.

\section{Results}
\label{sec:res}

We now present the results of our analysis. At first, in Section~\ref{subsec:SHMR}, we focus on the stellar-to-halo mass relation (SHMR) of our sample of haloes, comparing relations derived accounting for the triaxial shape of haloes and not. The following Sections explore the role of baryons and environment in the determination of structural properties of haloes. Section~\ref{subsec:cores} investigates the occurrence and efficiency of mechanisms related to the formation of density cores at the center of haloes hosting stars;  Section~\ref{subsec:barshapes} deals with the mechanism of central halo sphericization induced by baryons; Section~\ref{subsec:barshapes} presents a peculiar case of cusp regeneration in one of the haloes of the simulation; to conclude, Section~\ref{subsec:prolate} 
predominantly addresses the outer haloes regions and the processes that shape them. Finally, Section~\ref{subsec:lim} discusses whether and how physical processes not accounted for in the simulations could modify our results.

\subsection{The stellar-to-halo mass relation}
\label{subsec:SHMR}

Fig.~\ref{fig:SHMR} shows the stellar-to-halo mass relations (SHMRs) obtained matching virial masses from haloes that host stars with their corresponding stellar masses. The results are compared considering haloes whose masses are derived via a spherical binning scheme (left; $q=s=1$ in the previous sections) and when using triaxial shells (right), alongside analytic results extrapolated at our redshift/mass range from \cite{Moster2010}, \cite{Behroozi2013}, \cite{RodriguezPuebla2017}, and not extrapolated from \cite{Ma2019} and \cite{Rosdahl2022}. Stellar masses are always computed in triaxial bins. To help the comparison, in both cases we measured the slope ($m$) of the relation in the mass regime covered by the data by means of a linear fit in the $\log\Mvir-\log\Mst$ plane (grey dashed line)\footnote{The model's likelihood is 
\begin{equation}
    \ln\LL(\bxi) = -\frac{1}{2}\sum_{k=0}^{N}\biggl[\frac{\log\Mstk - m\log\Mvirk - q}{\sqrt{\Delta^2\log\Mstk+ m^2\Delta^2\log\Mvirk}}\biggr]^2,
\end{equation}
with $\log\Mstk$ and $\log\Mvirk$ the stellar and virial masses of the $k$-th halo, $\Delta\log\Mstk$ and $\Delta\log\Mvirk$ the corresponding average errors, and $m$ and $q$ the slope and height of the relation, respectively. We used an MCMC procedure similar to the one used in Sections~\ref{subsec:dm} and \ref{subsec:stars} to derive median parameters and uncertainties.}, inferring $m=1.40\pm0.03$ and $m=1.45\pm0.2$ when binning in spherical and triaxial shells, respectively.

\begin{figure*}
    \centering
    \includegraphics[width=1\hsize]{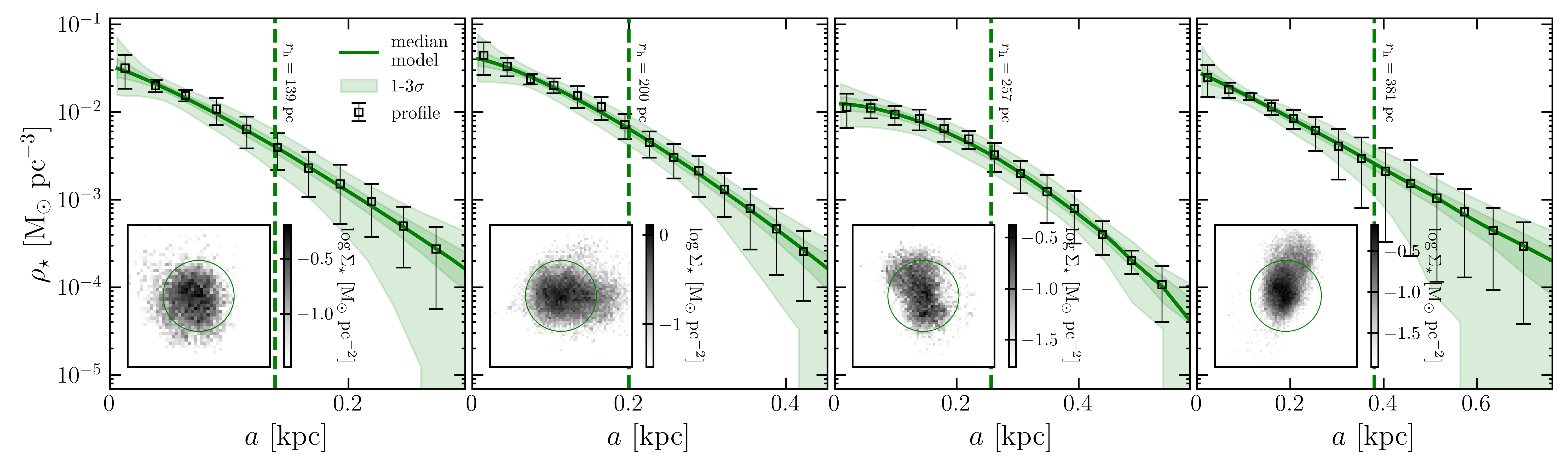}
    \includegraphics[width=1\hsize]{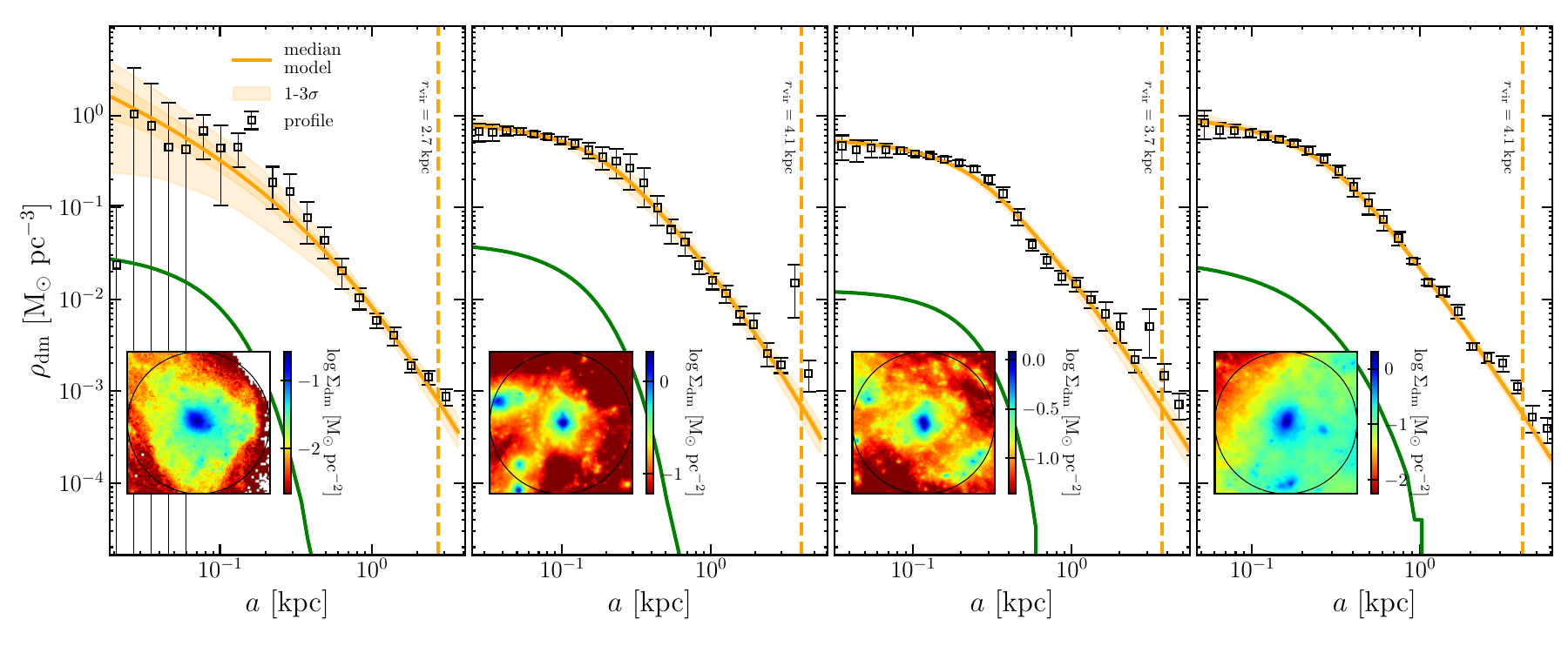}
    \includegraphics[width=1\hsize]{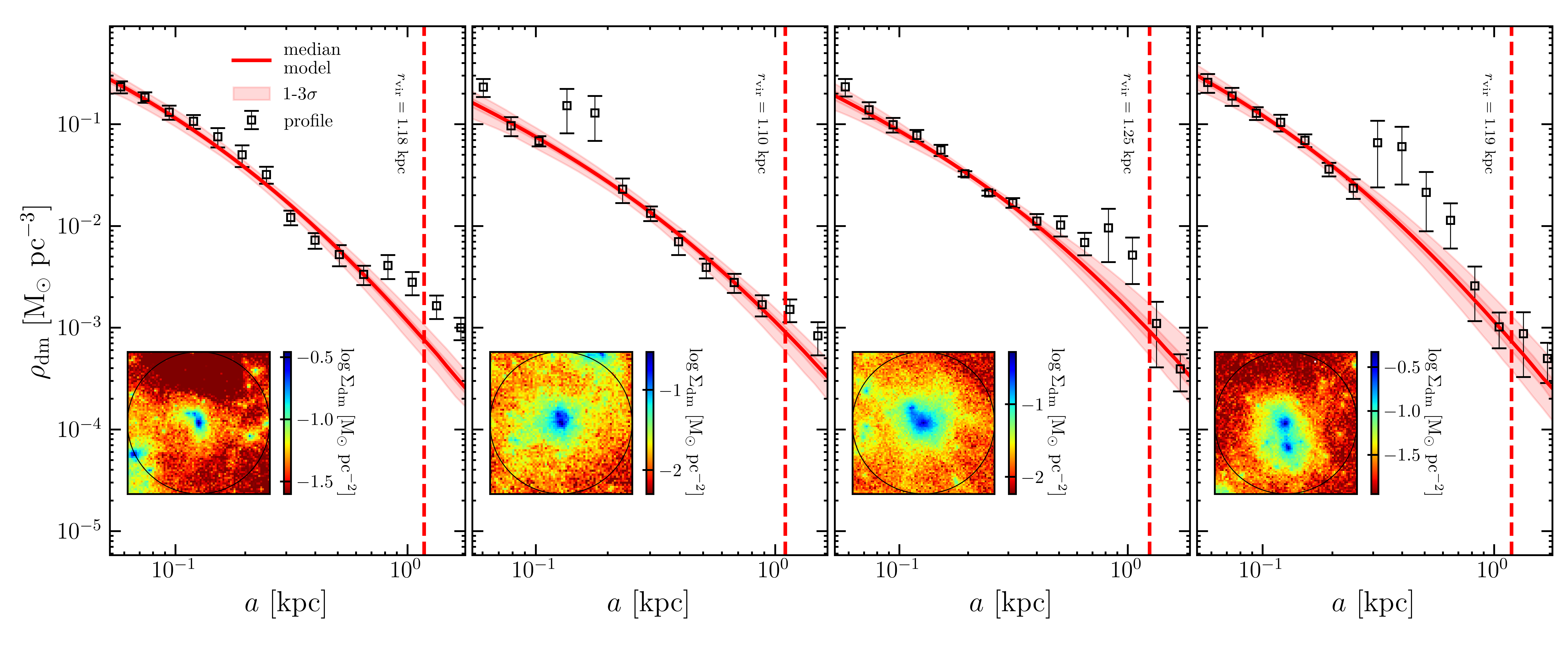}
    \caption{Top panels: binned stellar density distribution (squares with errorbars) superimposed to the median model (\ref{for:einasto}; green solid line) for a selection of four stellar systems. The dark and light bands show, respectively, the $1\sigma$ and $3\sigma$ uncertainties computed as described in Section \ref{subsec:stars}. The vertical dashed-green line marks the position of the stellar median half-mass radius. The small insets show the corresponding surface density distribution obtained assuming, as line-of-sight, the system $z$-axis. The green circle has an aperture given by the stellar median half-mass radius. Middle panels: same as the top panels, but showing the corresponding dark matter haloes (orange). Here, to help the comparison, we added the same median stellar densities shown in the top panels. The vertical dashed-orange line in the main panels and the black circle in the small insets mark the position of the halo virial radius. The binned density distributions are computed in triaxial bins of varying shape. Section \ref{subsec:dm} gives details on the fitting procedure. Bottom panels: same as the middle panels, but for a selection of four dark-matter haloes without a stellar counterpart (red). The position of the stellar systems and the corresponding dark-matter haloes in the middle panels are shown in the left panels of Fig.~\ref{fig:stdm}, while the position of the haloes in the bottom panels are shown in the right panels of Fig.~\ref{fig:stdm}.}
    \label{fig:stdmprof}
\end{figure*}

\begin{figure*}
    \centering
    \includegraphics[width=1\hsize]{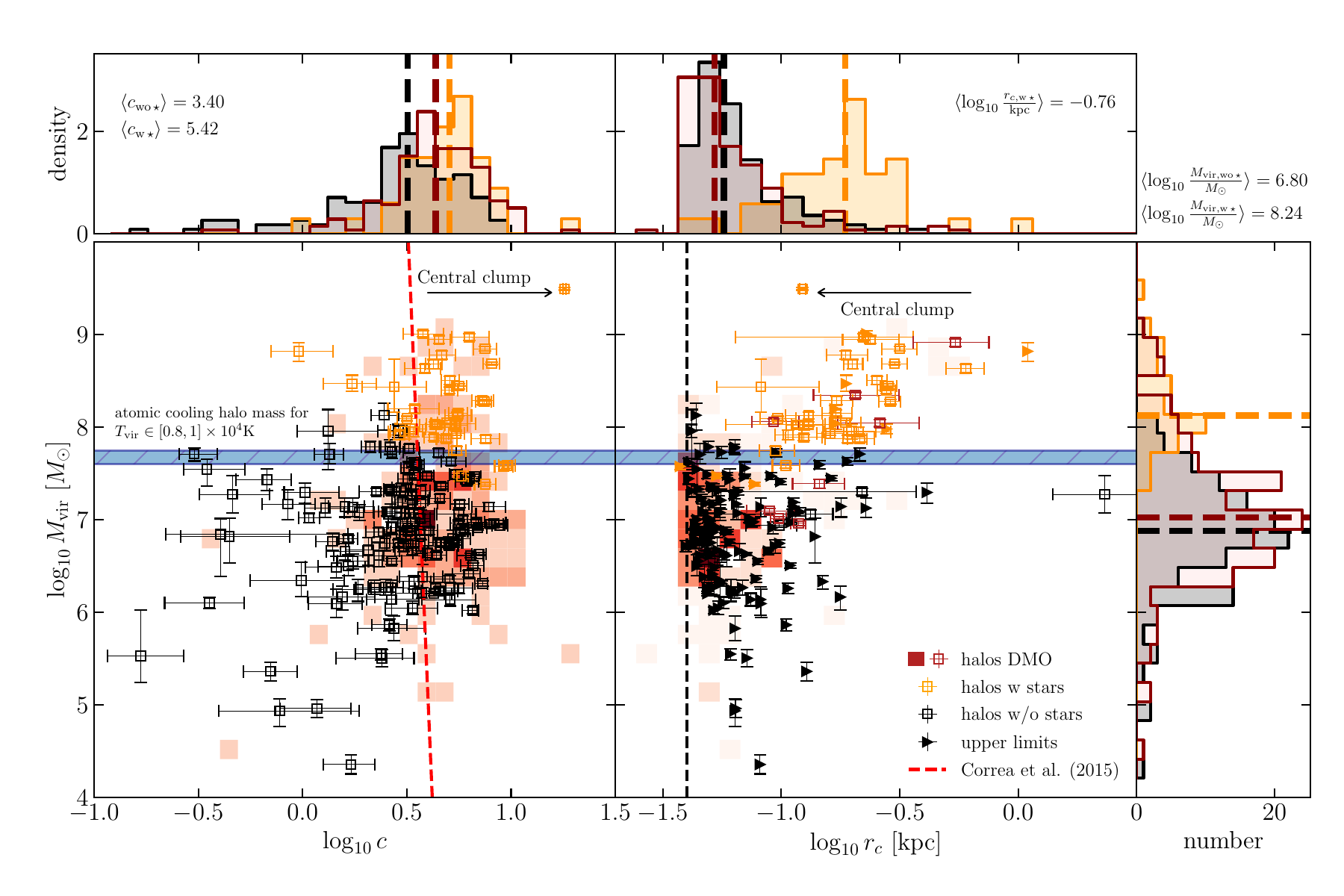}
    \caption{Halo virial mass $\log\Mvir$ as a function of the halo concentration $\log c$ (left panel; squares with errorbars) and as a function of the halo core radius $\log\rc$ (right panel). The orange squares correspond to haloes that have formed stars while the black squares to haloes that have not formed stars yet in the full physics simulation. The triangles in right-hand panel are, instead, upper limits on $\rc$. The small panels to the top and to the right show the corresponding one-dimensional distributions over $c$, $\Mvir$ and $\rc$, together with the median values of the distributions (dashed lines). Here, the distributions in the top panels are normalized to unity, while the one-dimensional distributions to the right show the number of haloes per mass bin. We show results from the DMO simulation as red two-dimensional distributions in the background of the $\log\Mvir-\log c$ and $\log\Mvir-\log\rc$ planes. Also, all values of $\rc$ that come from the fit in DMO simulation are upper limits. The horizontal blue band shows the mass range of an atomic cooling halo for virial temperatures $\Tvir\in[0.8,1]\times10^4\K$ (computed from equation 26 of \citet{Barkana2001} at $z=6.14$), the dashed red line in the left panel is the $\Mvir$ vs $c$ relation from \citet{Correa2015} at $z=6.14$, {and the dashed black line in the right panel marks the value of the smallest radial bin used to compute the density distributions of the starless and DMO haloes.}}
    \label{fig:logscrcMvir}
\end{figure*}

\begin{figure}
    \centering
    \includegraphics[width=.9\hsize]{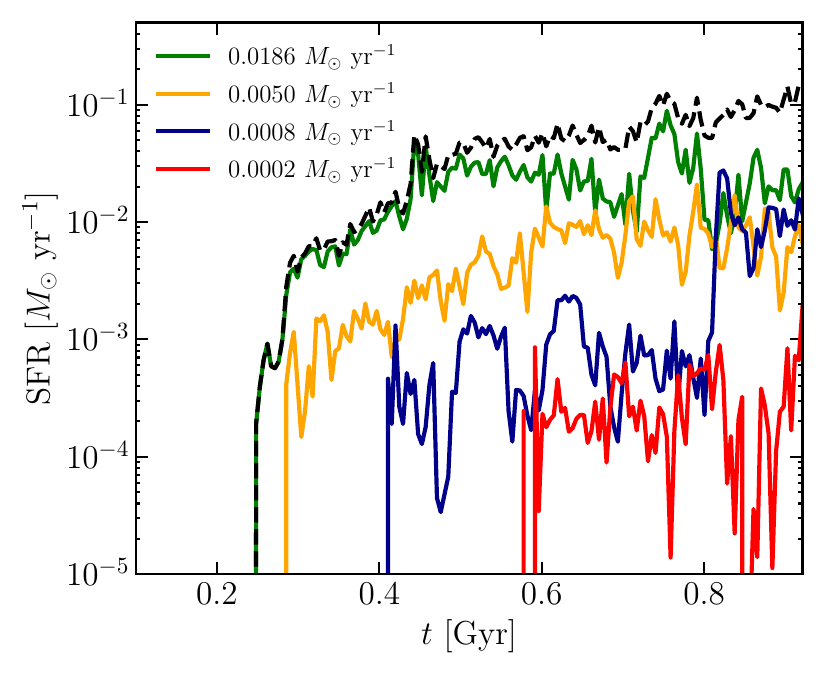}
    \caption{Selection of SFHs from 4 stellar systems in the simulation. The green curve is the SFH of the central galaxy while the black dashed line is the total SFH of the simulation. The legend reports the values of the median SFR computed over times of active star formation. The other stellar systems in the simulation have SFHs similar to the ones presented here.}
    \label{fig:SFH}
\end{figure}

At redshifts comparable to ours, \citet[shown by a red curve in Fig.~\ref{fig:SHMR}]{Ma2019} provides a statistically meaningful sample of haloes from a set of 34 high-resolution cosmological zoom-in simulations from the FIRE project (Feedback in Realistic Environments, \citealt{Hopkins2018}). \cite{Ma2019} consider haloes with virial mass between $10^{10.5}\Msun$ and $10^{12}\Msun$ for $5\le z\le12$ and, in their analysis, they do not find a significant evolution of the SHMR with redshift, measuring a slope $m=1.53$, consistent with ours within the intrinsic scatter of their sample ($\simeq 0.2$ dex). The offset between the \cite{Ma2019} relation and ours ranges from 0.57 at $\Mvir\simeq10^8\Msun$ to 0.46 at $\Mvir\simeq10^9\Msun$ in the triaxial binning scheme, and from 0.53 at $\Mvir\simeq10^8\Msun$ to 0.4 at $\Mvir\simeq10^9\Msun$ in the spherical binning scheme, in both cases with converging results towards higher masses. Our results are also consistent with \citet[shown by a green curve in Fig.~\ref{fig:SHMR}]{Rosdahl2022} who, analyzing galaxies in the \textsc{sphinx} suite of cosmological radiation hydrodynamical simulations resolving virial masses $\Mvir>7.5\times10^7\Msun$ at redshift $z=6$, derived a SHMR very similar to ours. The slope of the SHMR inferred by \cite{RodriguezPuebla2017} and \cite{Behroozi2013} is, respectively, $m=1.65$ and $m=2$, even less consistent with ours, even though their SHMR is extrapolated in the mass range covered by the stellar systems of our simulation. Recently, \cite{Ceverino2022} presented the analysis of galaxies from two sets of cosmological, zoom-in simulations from the VELA suite with different feedback models. Although they do not attempt to provide analytic fits to the SHMR, we are consistent with their relation at redshift $z=6$ at least for $\Mvir\simeq10^{10}\Msun$, the least massive halo masses covered by their simulations. Apart from \cite{Rosdahl2022}, all SHMRs from the literature shown in Fig.~\ref{fig:SHMR} imply lower stellar masses for a fixed halo mass than our SHMR, especially at the low mass end of the relation which is most sensitive to implementations of different stellar feedback. 

In the context of $N$-body numerical simulations, a precise measurement of the SHMR is essential for a realistic assignment of dark matter haloes to stellar clumps. We find that a single halo mass can differ by even a factor 1.6 with respect to its spherically binned analogous when accounting for its triaxial symmetry. However, the median effect on the SHMR is negligible since the median ratio between 'spherical' and 'triaxial' virial masses for haloes in our sample is unitary. Despite haloes are always expected to depart from sphericity, especially in the outer regions where the vast majority of the mass resides \citep{VeraCiro2011}, our analysis, that only focuses on $z=6.14$, suggests that we may expect similar results (i.e. negligible effect on the SHMR when account for triaxiality) at all redshifts.

\subsection{Baryon-driven flattening of cusps in dark matter haloes}
\label{subsec:cores}

We now give details on the properties of the stellar systems formed in the simulation and we consider the effects that star formation has on halo profiles. In the top row of panels in Fig.~\ref{fig:stdmprof}, we show the binned stellar density distribution superimposed to that of the median model of equation (\ref{for:einasto}) for a selection of 4 stellar systems. The small insets show the corresponding surface density maps projected along the same line-of-sight as in the top panels of Fig.~\ref{fig:stdm} where, to facilitate the comparison, they have also been marked by orange circles. The middle panels show, instead, the profiles and surface density maps of the dark matter haloes in which the clumps in the top panels are embedded, while the bottom panels are density distributions of a selection of four dark matter haloes that have not formed stars yet at $z=6.14$ (shown by black circles in the right panels of Fig.~\ref{fig:stdm}). On average, we find that dark matter dominates over stars at all radii. As a reference, the median ratio between dark matter and stellar densities at the stellar half-mass radius is $\simeq100$, with a larger dispersion towards higher ratios than lower ratios. The median stellar half-mass radius is about $\simeq200\pc$, with the smallest clump having a half-mass radius of $90\pc$. The shape and size of the stellar distributions depend on the stellar clump's total mass: massive systems ($\Mst\geq10^5\Msun$) have, on average, index $n\simeq0.8$ (equation \ref{for:einasto}) and larger half-mass radii, while the low mass ones ($\Mst\leq10^5\Msun$) have a median $n>1$, a wide marginalized posterior distribution over $n$, and have lower half-mass radii. On average, the most massive stellar systems in the simulation are also more concentrated and less diffuse than the least massive ones. In terms of mass, size, and dark matter content, the newly formed stellar systems resemble dark-matter dominated dwarf galaxies (see also Fig. 12 of \citealt{Calura2022}).

As for the dark matter haloes, a quantitative insight on their properties is given in Fig.~\ref{fig:logscrcMvir}. Here, the left panel shows the halo virial masses against the halo concentrations, for all haloes identified, following the procedure in Section~\ref{sec:id}, in both the full physics and DMO simulations. In case of the full physics simulation, haloes are differentiated by color, with haloes that have formed stars at $z=6.14$ being shown in orange and haloes that have not formed stars in black. For clarity, haloes in the DMO simulation are shown in the background as two-dimensional distributions. The distinction between families of haloes with and without stars is sharp and falls around $\Mvir\simeq5\times10^7\Msun$, which corresponds to the virial mass of an atomic cooling halo (computed by inverting equation 26 in \citealt{Barkana2001} and shown in Fig.~\ref{fig:logscrcMvir} by a blue band assuming $\Tvir=8000\K$ and $\Tvir=10000\K$). As dark matter structures increase in mass through the accretion of smaller systems, gas is heated up to the halo virial temperature. Since the simulation lacks molecular hydrogen cooling and metallicity is initialised to zero at the beginning of the simulation, gas only cools via atomic cooling that is effective down to $T\simeq 10^4\K$. When the halo virial temperature becomes greater than this value, gas starts to be confined by the potential well of the halo, and it can efficiently trigger star formation. It is then evident, and it follows throughout the rest of this paper, that a distinction between haloes that have formed and not formed stars naturally implies a selection in halo mass, with $\log\Mvir/\Msun\gtrsim7.6$ indicating the former and $\log\Mvir/\Msun\lesssim7.6$ the latter. We also note that star forming haloes are more concentrated than starless haloes in the full physics simulation. This may be due to adiabatic contraction (e.g. \citealt{Blumenthal1986,Gnedin2004}), which is more efficient in haloes above the atomic cooling limit, since these haloes can efficiently cool gas and accumulate baryons at the center.

The large right-hand panel of Fig.~\ref{fig:logscrcMvir} shows halo virial masses as a function of the median core radius $\rc$ of the coreNFW model used in the fitting procedure of Section~\ref{subsec:dm}. The plot indicates that when haloes reach the $\log\Mvir/\Msun\sim7.6$ mass threshold, the ignition of star formation (and related phenomena such as winds, SN feedback, gas cooling and condensation) has a significant impact on the redistribution of dark matter particles within the central regions of the haloes, erasing the classical central $r^{-1}$ cusp in favor of a core of constant density. This effect of cusp flattening is unseen in the DMO simulation, where haloes maintain their original cusp in the central regions. Indeed, we point out that all measurements of $\rc$ shown in Fig.~\ref{fig:logscrcMvir} obtained for starless haloes and haloes in the DMO simulation resulting from the fitting procedure of Section~\ref{subsec:dm} are upper limits. Only star-forming haloes have a well defined lower limit on $\rc$ (see also Fig.~\ref{fig:stdmprof}). 
When evaluating the binned dark-matter density distributions, in case of starless haloes and haloes in the DMO simulation, we set the smallest radial bin of the profile to $40\pc$ (vertical dashed line in the right hand panel of Fig.~\ref{fig:logscrcMvir}), a value smaller than the minimum stellar half-mass radius measured among the population of stellar systems. In case of haloes with stars, instead, we set this number to 10\% of the corresponding stellar half-mass radius. This ensures us to have a good inference power on sizes comparable to that of stars, crucial to resolve the effects of baryons and of the growth of a stellar component in the center of haloes. In this respect, we note that the upper limits found on $\rc$ are always of the order of this $40\pc$ minimum resolution.

The formation of cores at the center of dark matter haloes has been long debated \citep{Bullock2017,DelPopolo2022}. The tension, named the 'core/cusp problem', comes from observations of present-day dwarf galaxies, in which a constant dark matter density distribution is inferred from measures of central rotation velocities \citep{McGaugh1998,deBlok2003,Gentile2007,Salucci2007,Oh2015}, whereas dark matter haloes from early $N$-body, DMO simulations preferentially form cusps \citep{Dubinski1991,NavarroEF1996}. Cores of nearly constant density can be formed, however: i) naturally in a cosmological context with other cosmological models than the classical $\Lambda$CDM paradigm, invoking, for instance, exotic dark matter particles \citep{Nadler2021,Hui2017}; ii) via interactions with baryons missed or usually not accounted for or not properly described in simulations. Focusing on ii), \cite{Pontzen2012} and \cite{Teyssier2013} demonstrated that sudden and oscillating changes in the gravitational potential driven by SN explosions, and the consequent shift towards equilibrium via violent relaxation \citep{LyndenBell1967}, can efficiently sweep low angular momentum gas at the centers of haloes, becoming a viable process to transfer energy to dark matter, heating it up and erasing the cusp over less than a cosmic time. The mechanism is efficient and produces relatively extended cores also when a small fraction of baryons at the center of haloes forms stars, even with low SN efficiency \citep{Maxwell2015}, as long as the star formation rate (SFR) remains bursty. \cite{NipotiBinney2015} showed that the flattening of the cusp can happen even before the onset of star formation. As the gas is accumulated at the center of the halo in a disc, it eventually becomes denser than the local dark matter density and unstable against gravity. When the disc fragments, the remaining gas clumps shift towards the halo center and transfer energy to the dark matter cusp via dynamical friction. This process erases the cusp on short timescales. 

Thanks to the high spatial resolution of our simulation and to the possibility to resolve stellar feedback from individual massive stars at very high redshift, we are able to witness the interplay between baryons and dark matter  that shapes low-mass young haloes in a fully cosmological context. Although we will dedicate a detailed study to the properties of the stellar clumps detected in the simulation in a second paper (Pascale et al. in preparation), all stellar clumps found here present a very irregular and discontinuous star formation. We show, for a qualitative comparison, the SFRs of four stellar clumps in Fig.~\ref{fig:SFH} (the green line is the central clump). In all cases the SFR is bursty and discontinuous, with variations from the median SFR of even a factor of 10, in agreement with requirements for core formation from previous works. In our case, we believe that dynamical friction of gas may have contributed to the early flattening of the cusp, since several haloes around the mass threshold $\log\Mvir/\Msun\simeq7.6$, but without stars, have large upper limits over $\rc$. However, only for haloes that have formed stars there is evidence that the dark matter halo is cored, indicating that the onset of stellar feedback from winds and supernovae explosions is predominantly efficient. In Fig.~\ref{fig:logscrcMvir}, it can be observed that, for star-forming haloes, there is a distinct trend of an increasing halo core size with halo mass. As \cite{Calura2022} have demonstrated, there is a positive correlation between the stellar mass of the clumps within the simulation and their half-mass radius $\rh$ (see Fig. 12 of \citealt{Calura2022}). This implies that there is also a positive correlation between the halo core size ($\rc$) and the size of the corresponding stellar clumps, and further suggests that baryonic matter directly impacts the host haloes on scales smaller than a few $\rh$.


A similar analysis has been conducted by \cite{Fitts2017} who studied isolated galaxies with $\Mvir\simeq10^{10}\Msun$ at $z=0$ from a set of 15 high resolution, cosmological simulations within the FIRE project. In the case of \cite{Fitts2017}, the cusp flattening is measured in all those haloes hosting galaxies with stellar mass (at $z=0$) higher than $\simeq2\times10^6\Msun$, corresponding to $\Mst/\Mvir = 2\times10^{-4}$. In our case, instead, core formation happens also in the least massive galaxies of the sample, with $\Mst\simeq10^{4-5}\Msun$, corresponding to $\Mst/\Mvir \simeq10^{-3}-10^{-4}$. Although our results differ from those of \cite{Fitts2017}, it is important to consider that the two analyses differ in several significant aspects. First of all, this comparison involves galaxies at very different redshifts, which may reflect completely different environments and physical conditions for galaxy evolution (e.g., at $z=6.14$ the galaxies in our sample are still star-forming while most of the low-mass galaxies in \citet{Fitts2017} are quenched); also, the two sets of simulations have very different resolution and feedback prescriptions, e.g. in FIRE, the initial metallicity is $10^{-4}$ solar, hence gas cools down and fragments more easily at early times, forming stars in smaller haloes, whereas in our case we are limited to atomic cooling haloes \citep{Calura2022}. Despite these differences and the exact mass threshold for core formation, \cite{Fitts2017} also report a distinct increasing trend of the halo core size with the stellar mass, with cores more extended in more massive galaxies, and always of the order of the stellar half-mass radius.

Other processes that can influence the formation and evolution of cores in haloes are interactions and mergers with smaller systems, resulting in the so called cusp regeneration \citep{Laporte2015,Orkney2021}. We will delve further into this topic by examining the merger history of the central, massive clump in the simulation in Section~\ref{subsec:ccentral}.


\subsection{Effect of baryons on the central dark matter halo shape}
\label{subsec:barshapes}

\begin{figure*}
    \centering
    \includegraphics[width=0.49\hsize]{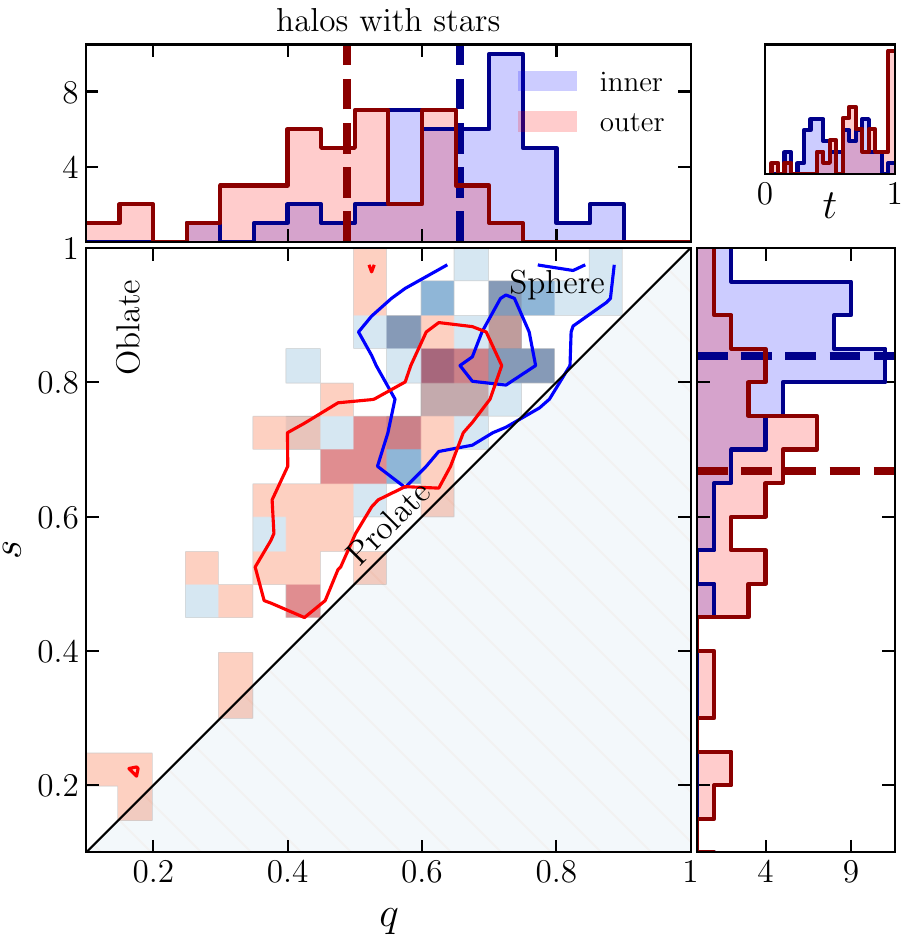}
    \includegraphics[width=0.49\hsize]{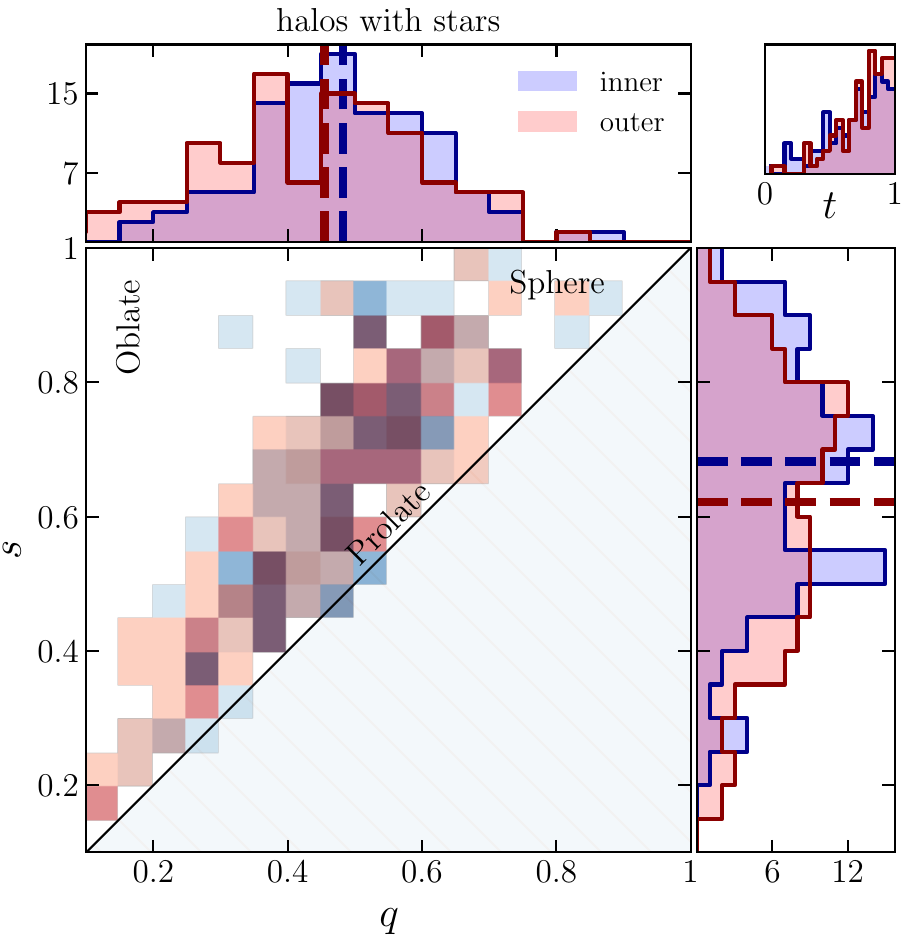}
    \caption{Left panels: two-dimensional distributions (bottom left panel) of intermediate-to-major ($s$) and minor-to-major ($q$) axes of the dark matter haloes that have formed stars within the simulation at redshift $z=6.14$, alongside the corresponding one-dimensional, marginalized distributions $q$ (top left panel), $s$ (bottom right panel) and triaxiality $t$ (equation~\ref{for:tri}, top right panel). Blue colors refer to inner ($[0,0.25]\rvir$) while red colors to outer halo shapes ($[0.75,1]\rvir$). Details on the procedure used to determine the halo shapes are given in Section~\ref{subsec:stars}. The diagonal in the two-dimensional distribution marks the allowed portion of the $s-q$ space ($q\le s\le 1$). The vertical(/horizontal) dashed line in the top (right) panel shows the median $q$ ($s$). Right panels: same as the left panels but for the haloes that do not have formed stars.}
    \label{fig:shape}
\end{figure*}

\begin{figure*}
    \centering
\includegraphics[width=1\hsize]{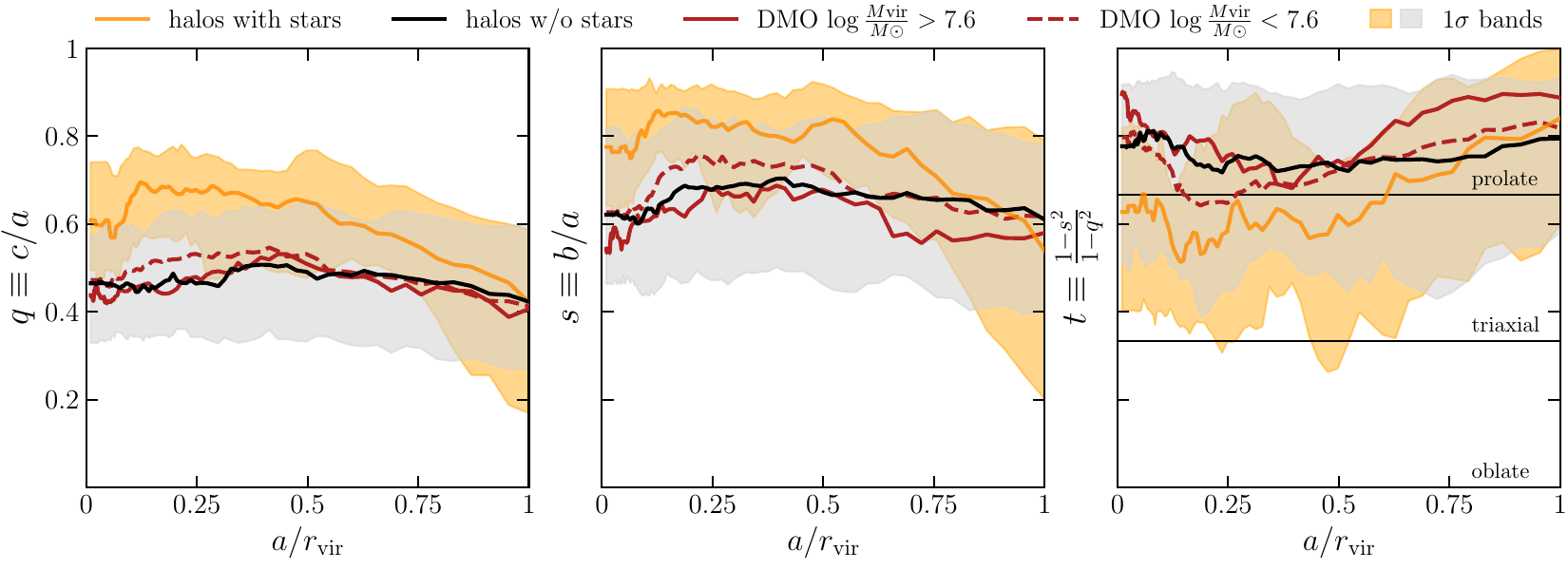}
    \caption{Left panel: median minor-to-major axis ratio as a function of the semi-major axis normalized to the halo virial radius for haloes that have formed (orange solid line) and have not formed stars (black solid line) in the main simulation, and haloes above (red solid line) and below (red dashed line) the mass cut $\log\Mvir/\Msun\simeq7.6$ in the DMO simulation. The band is the region between the 16th and 84th percentiles of the one-dimensional $q$ distribution for any fixed $a/\rvir$. For clarity, we only show the band for the star-forming and non-star forming haloes. The profiles obtained for haloes in the DMO simulation have, however, a very similar dispersion to that of the non-star-forming haloes. Middle panel: same as the left panel, but showing the median intermediate-to-major axis $s$ profile. Right panel: same as the left panel, but showing the median triaxiality parameter $t$ profile (equation \ref{for:tri}). The single halo profiles have been interpolated with a cubic spline and re-sampled to compute the corresponding distributions.}
    \label{fig:shapeprof}
\end{figure*}

We now focus on studying the inner shape of our sample of haloes, with particular attention on the mechanisms determined by baryons. The left panel of Fig.~\ref{fig:shape} shows the two dimensional distribution (bottom-left) of intermediate-to-major ($s$) and minor-to-major axes ($q$), the corresponding one dimensional, marginalized distributions of $s$ (bottom-right), $q$ (top-left) and of the triaxiality parameter $t$ (top-right) considering the dark matter haloes that host stars. The triaxiality parameter is  \citep{Franx1991}
\begin{equation}\label{for:tri}
    t \equiv \frac{1-s^2}{1-q^2}, 
\end{equation}
and it measures the prolateness or oblateness of an ellipsoid. As a convention, $t<\frac{1}{3}$ indicates oblate systems ($q<s\simeq1$), $\frac{2}{3}<t<1$ prolate systems ($q\lesssim s$), while to $\frac{1}{3}<t<\frac{2}{3}$ corresponds to triaxial systems. We also define the innermost region of a halo as the triaxial ellipsoid with semi-major axis that extends up $0.25\rvir$, while the outermost region corresponds to the radial range $[0.75,1]\rvir$. For the purposes of Fig.~\ref{fig:shape}, for each halo we have recomputed $q$ and $s$ in these bins adopting as $\rvir$ the median value resulting from the fitting procedure of Section~\ref{subsec:dm}. We will, therefore, refer to the halo \textit{inner shape} as the shape computed within $0.25\rvir$, and as halo \textit{outer shape} the one computed within $[0.75,1]\rvir$. In Fig.~\ref{fig:shape} we have marked with different colors distributions of inner (red) and outer shapes (blue) to empathize the two different behaviors. 

It is particularly appreciable that haloes that have formed stars ($\log\Mvir/\Msun\gtrsim7.6$) are more spherical in the central regions while they become more elongated and prolate in the outer parts, with a median $\qmean$ and $\smean$ decreasing from $0.66_{-0.13}^{+0.11}$ to $0.49_{-0.17}^{+0.14}$, and from $0.84_{-0.12}^{+0.07}$ to $0.67_{-0.31}^{+0.14}$, respectively. In a similar manner, the pattern is also evident when examining the one-dimensional $t$ distribution: while only 33\% of haloes have triaxiality $t>\frac{2}{3}$ in the inner parts, the percentage increases to 68\% outwards, with a median $\tmean$ ranging from $0.58_{-0.23}^{+0.20}$ (inner) to $0.73_{-0.20}^{+0.24}$ (outer), and an overall distribution skewed towards large $t$. The right hand set of panels of Fig.~\ref{fig:shape} shows the same distributions of $q$, $s$ and $t$ as in the left panels, but considering haloes without stars. In this case, when stars are absent, the picture is completely different: haloes are arranged along the bisector of the two-dimensional $q-s$ space and thus are prolate, with no significant difference between the inner and outer regions.

Given that the sample of haloes identified in the simulation encompasses a very broad range of masses ($\log\Mvir/\Msun\simeq5.5-9$), the separation between star-forming and non-star-forming haloes at $\log\Mvir/\Msun\simeq7.6$ necessitates the comparison of haloes with considerably different masses. This large difference raises the question of whether the distinct behaviors of inner shapes are truly attributable to the presence of baryons, or if it is biased by the fact that the two classes of haloes represent dynamically distinct objects, with the smaller ones being younger and potentially having a very different mass assembly history. In Fig.~\ref{fig:shapeprof} we show the median $s$, $q$ and $t$ as a function of the haloes' semi-major axis normalized to the virial radius. For a fairer comparison, in this case we have also included the median shapes profiles computed for the dark matter haloes identified in the DMO simulation. Different types of haloes are labelled by different colors: orange and black are star-forming and non-star-forming haloes in the principal simulation, respectively, while in red we show the profiles derived for haloes in the DMO simulation. In this latter case, we have applied the same mass cut $\log\Mvir/\Msun\simeq7.6$ showing different profiles for different mass ranges. It is now much more evident that, on average, haloes that do not host stars and all haloes in the DMO simulation have very similar $q$, $s$ and $t$ profiles that are manteined approximately constant throughout their full radial extent. Haloes with stars, on the contrary, populate very different regions of the plot and are the only ones that are, in their centers, much more spherical than all the others. Both median inner $\smean$ and $\qmean$ of haloes without stars and in the DMO simulation are approximately 0.2 dex smaller than the same values measured from massive, star-forming haloes, with triaxiality parameter distribution being strongly shifted towards unity (right-top panel of the Fig.~\ref{fig:shape} and right-hand panel of Fig.~\ref{fig:shapeprof}).


\begin{figure*}
    \centering
    \includegraphics[width=1.\hsize]{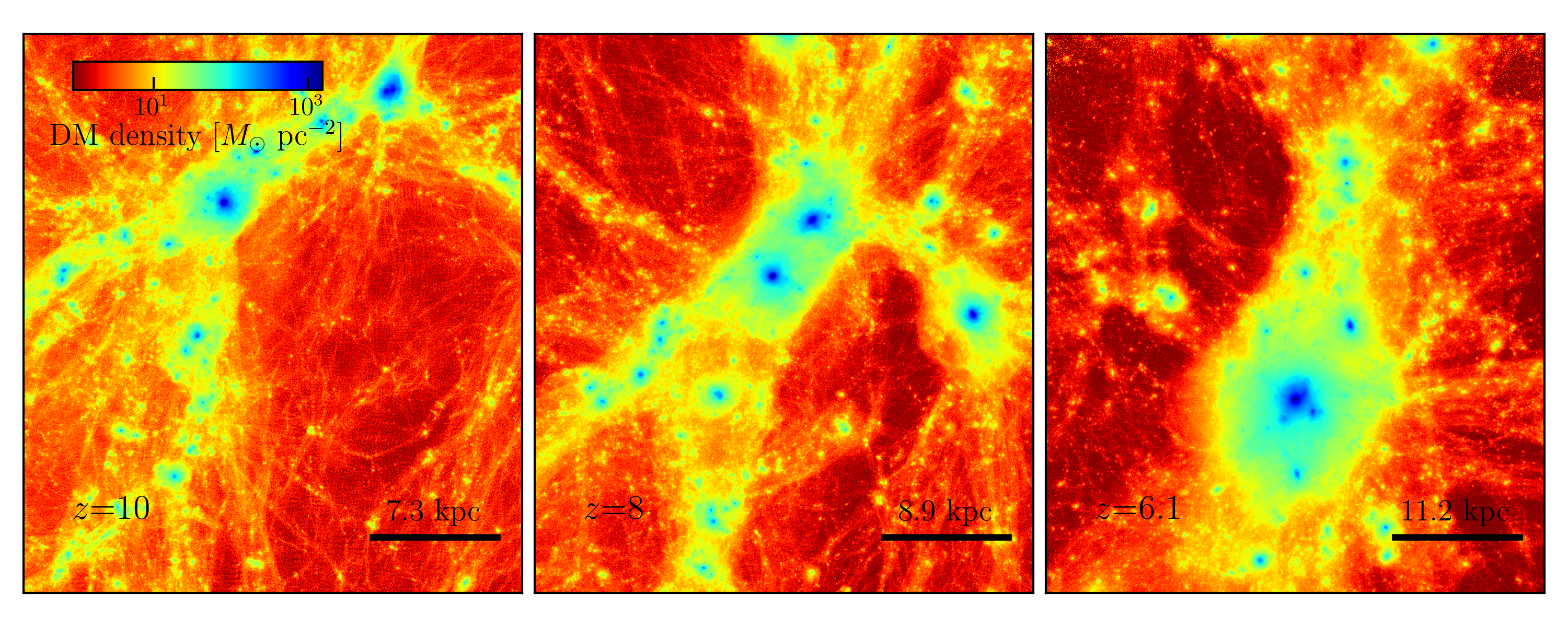}
    \includegraphics[width=1.\hsize]{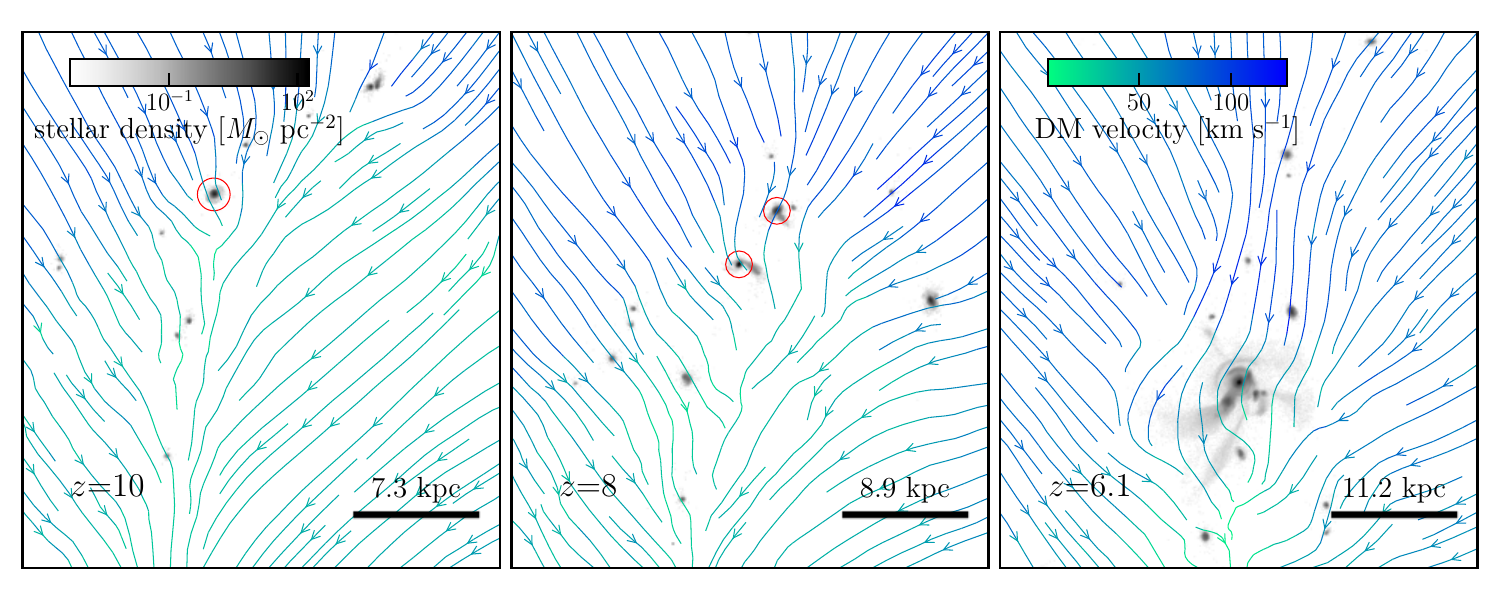}
    \caption{Top panels: series of dark matter surface density maps from different snapshots of the full physics simulation. The maps show a zoom-in view of the region around the central and most massive dark matter halo of the simulation at $z=10$, $z=8$, $z=6.14$, respectively from left to right. Blue colors correspond to high density regions while red colors to low density regions. Bottom panels: same as the top panels but showing the stellar density maps. In the bottom panels we have superimposed the dark matter streamline velocity field parallel to the plane of the image. In the bottom-left and bottom-middle panels we mark the haloes that have been included in the analysis with a red circle.}
    \label{fig:simz}
\end{figure*}

\begin{figure*}
    \centering
    \includegraphics[width=1\hsize]{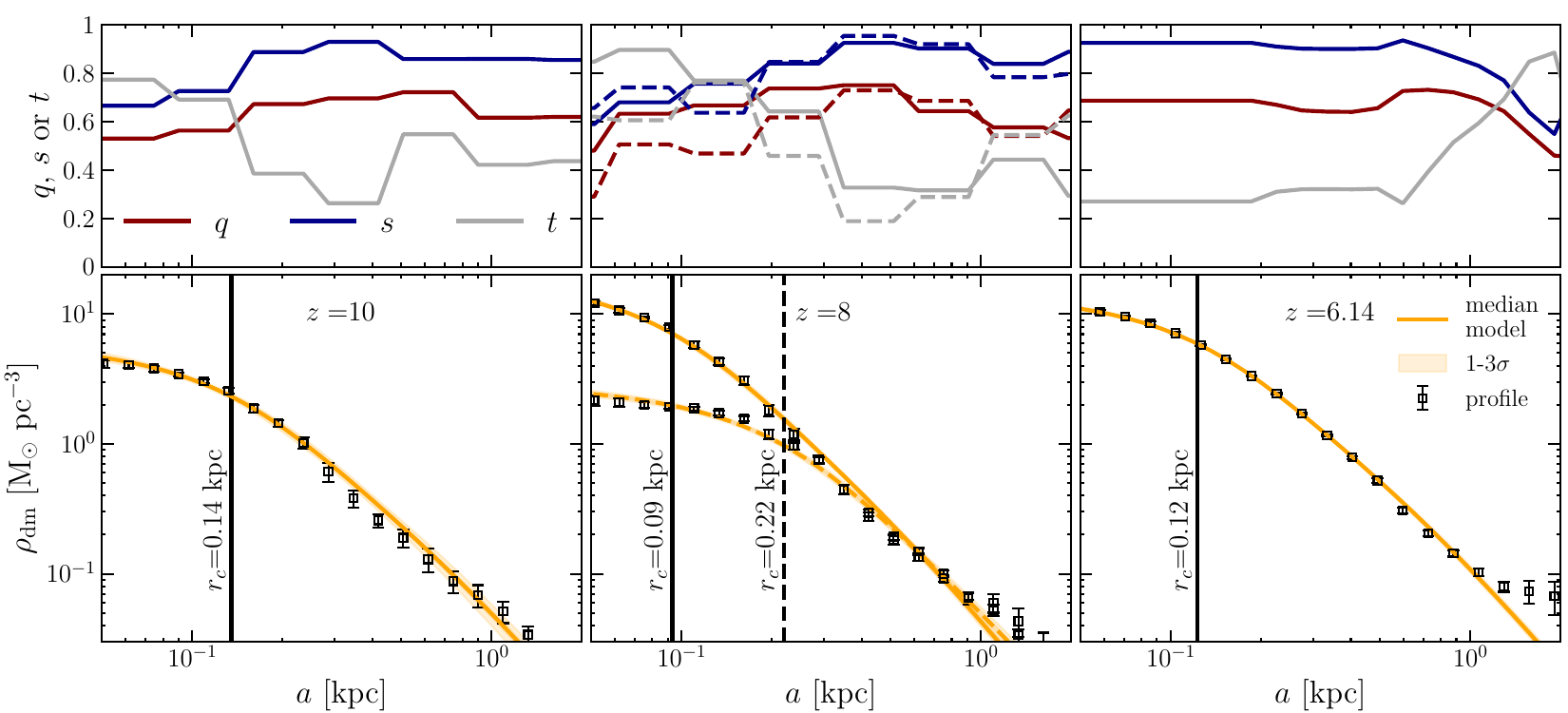}
    \caption{Top panels: minor-to-major axis $q$ (solid red line), intermediate-to-major axis $s$ (solid blue line) and triaxiality parameter $t$ (solid grey line) as a function of the semi-major axis $a$ for the progenitors of the central halo for decreasing redshifts (from left to right). In the top middle panel, to distinguish between the two haloes at $z=8$ we use different linestyles. Bottom panels: binned dark-matter density distributions (squares with errorbars) superimposed to that of the median coreNFW model for the progenitors of the central halo in the same redshifts as in the top panels. The dark and light orange bands show, respectively, the $1\sigma$ and $3\sigma$ uncertainties. The vertical solid and dashed-black lines mark the positions of the median halo core radius.}
    \label{fig:centralqst}
\end{figure*}

Analogous to the cusp-flattening discussed in the previous Section, processes related to star formation are the primarily culprits for reshaping the inner haloes. \cite{Blumenthal1986} argued that the infall of cool gas increases the central density of dark matter haloes, due to adiabatic contraction, leading to a change of the orbital distribution in the central halo that ultimately reshapes the inner halo towards a less prolate/rounder configuration \citep{Dubinski1994,Debattista2008}. This result has later been supported by several studies focusing, via hydrodynamical $N$-body simulations, on the present-day shape of dark matter haloes and its dependence on the halo baryonic content. Comparing adiabatic (i.e. without cooling) and full physics (i.e. with star formation, supernovae, metal enrichment, cooling and UV ionizing background) simulations of galaxy groups and isolated, MW-like,  galaxies, \cite{Kazantzidis2004} found that baryonic dissipative processes make the central dark-matter haloes rounder, with a decreasing trend towards the outer parts. In a similar way, analyzing haloes from the NIHAO suite of high-resolution cosmological simulations, \cite{Butsky2016} found rounder halo inner shapes when full baryonic physics are accounted for, with an average minor-to-major axis ratio $q\simeq0.8$ in galaxies with masses comparable to that of the MW. However, especially at the low mass-end of their halo sample ($<10^{11}\Msun$), they do not find strong differences with respect to DMO simulations, in contrast with our results. Studying haloes from Illustris TNG100 and TNG50 simulations and galaxies from EAGLE and NIHAO at $z=0$, and comparing results with analogous DMO simulations, in agreement with past results, \cite{Chua2022} also found rounder shapes at the center of haloes in the virial mass range $10^{10–14}\Msun$, with a less pronounced effect for less massive haloes. They also explored variations of halo shapes resulting from variations of stellar feedback models implemented (i.e. strength of stellar wind and BHs) via smaller box hydrodynamical simulations. Although they find that different feedback prescriptions have a different degree of impact on halo shapes, with stronger/faster galactic winds producing the least spherical haloes, all feedback models they explore produce results significantly different from those of similar haloes in DMO simulations, with inner shapes anyway rounder and with shape profiles close to one another. \cite{Bryan2013} also found that baryons have a non negligible effect on inner halo shapes for a wide variety of halo masses (galaxies to clusters) up to redshift $z=2$ looking at haloes extracted from the OWLS suite of cosmological simulations (OverWhelmingly Large Simulations). Very similar results were obtained by \cite{Prada2019} looking at MW-like galaxies in the AURIGA simulations, and \cite{Cataldi2021} with the Fenix and EAGLE cosmological simulations, also finding a dependence of halo shapes on galaxy morphology. 

We not only confirm that baryons make dark matter haloes rounder in the central parts, but we also provide quantitative evidence that the effect occurs as soon as haloes are illuminated by the first stars, a process that, in principle, sets in at very high redshift. In this respect, we extend previous results from studies that primarily focus on the present-day shape of massive haloes to theirs high redshift, low-mass halo progenitors.

\subsection{Sphericization and cusp regeneration}
\label{subsec:ccentral}

We select two additional snapshots at redshifts $z=10$ and $z=8$, when the Universe is $0.48\Gyr$ and $0.65\Gyr$ old, respectively, and explore to what extent structural properties of haloes such as shape and density distribution depend on the halo merger history, on redshift and on the growth of the stellar component in its center. For the sake of simplicity, we restrict the analysis to the most massive halo of the simulation at $z=6.14$, which guarantees, even at these high reshifts, sufficient sampling by both dark matter and stellar particles. 

The sequence of panels in the upper row of Fig.~\ref{fig:simz} shows a zoom-in view of the central region of the simulation box. The panels follow the evolution and growth of the central halo and the dark-matter filaments connecting to it. The bottom row of panels shows, instead, the corresponding stellar surface density maps superimposed with dark matter streaming velocity maps in the plane of the image. The central halo at $z=6.14$ is the result of a major merger between two systems with similar mass in the redshift interval 8-6.14. We identify the haloes and compute their centers, density distributions and shapes following the same procedures as described in Section~\ref{subsec:dm}. At $z=8$, the progenitors have virial masses $\simeq8.4\times10^8\Msun$ and $\simeq10^9\Msun$, which implies, approximately, a 1:1 merger. At $z=10$ we focus only on one of the two progenitors, shown to the left of the panel of Fig.~\ref{fig:simz} by a red circle, whose structure is sufficiently regular to allow us to derive its properties with confidence. At $z=10$, the halo virial mass is $8.7\times10^8\Msun$, while the central stellar system has built 16\% of its total stellar mass at $z=6.14$, which increases to 50\% at $z=8$, meaning that, subsequently to the merger, the central halo builds the vast majority of its stellar mass (see the green line in Fig.~\ref{fig:SFH} which represents the SFH of the central galaxy).

As mentioned in Section~\ref{subsec:cores}, the central galaxy has one of the smallest core size at redshift $z=6.14$ ($\rc\simeq0.12\kpc$) despite having the largest mass among all haloes examined (see Fig.~\ref{fig:logscrcMvir}), in contrast to the other haloes. The median core sizes of the progenitors at redshift $z=8$ are $\rc\simeq0.9\kpc$ and $\rc\simeq0.22\kpc$, in one case smaller to the one at $z=6.14$, in the other case larger. At $z=10$, instead, $\rc\simeq0.14\kpc$, still larger. This means that, as the central halo grows in time, it reduces the size of its density core while increasing the extent of its stellar component. In other words, its cusp is regenerating. This cusp regeneration is more evident in the bottom panels of Fig.~\ref{fig:centralqst}, where we show the dark matter density profiles of the central halo at redshift $z=6.14$ and of its progenitors. 

\begin{figure}
    \centering
    \includegraphics[width=1.\hsize]{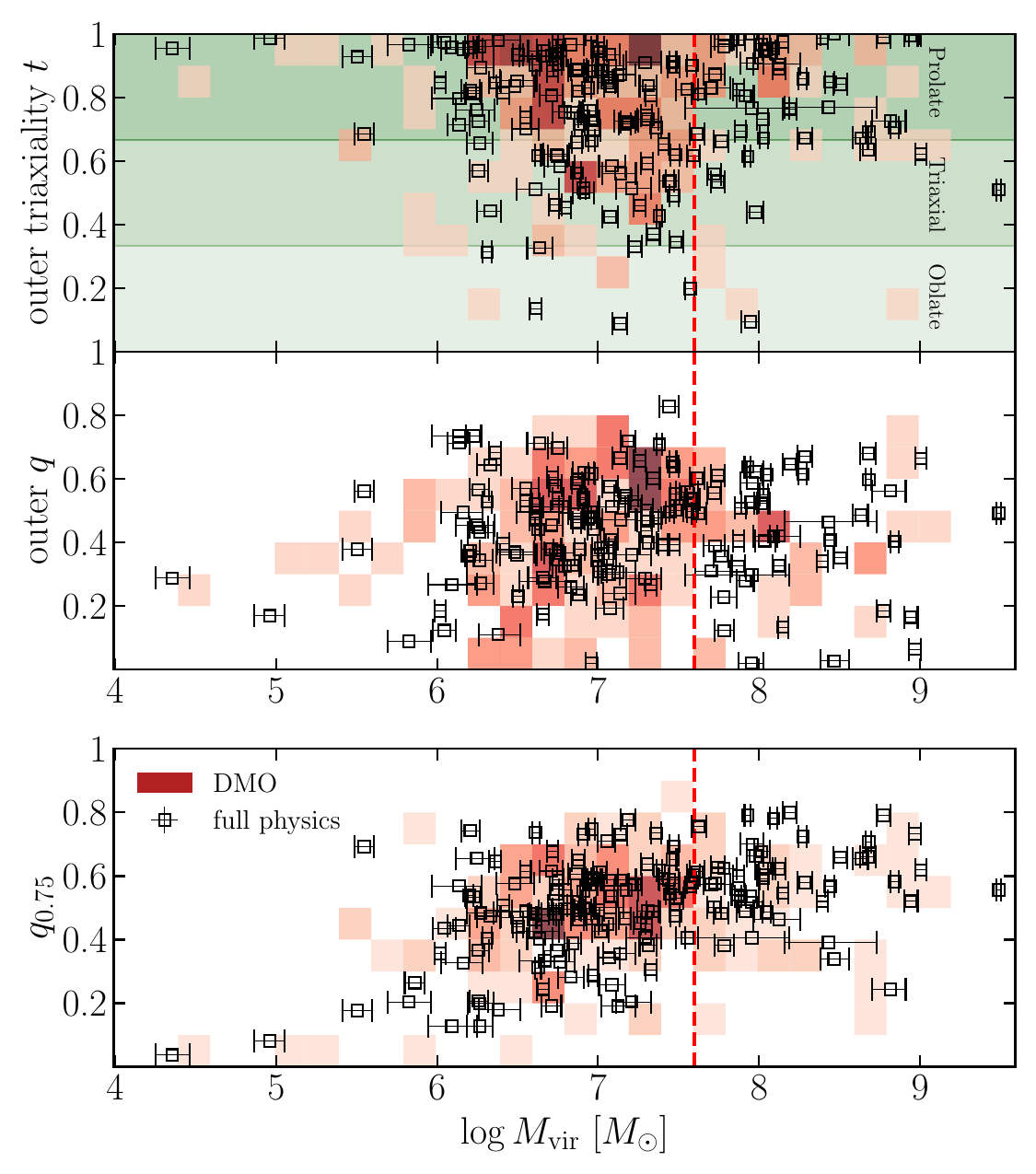}
    \caption{Top panel: outer triaxiality parameter against halo virial mass for all haloes identified within the simulation (points with errorbars) and haloes in the DMO simulation (red distribution). Middle panel: same as the top panel but showing the outer minor to major axis $q$. Bottom panel: same as the top panel but showing  $q_{0.75}$, i.e. the minor to major axis computed considering dark matter particle within $[0.5,0.75]\rvir$. The vertical red dashed line shows the star formation threshold mass $\log\Mvir/\Msun\simeq7.6$.}
    \label{fig:tvsMvir}
\end{figure}

\begin{figure*}
    \centering
    \includegraphics[width=1\hsize]{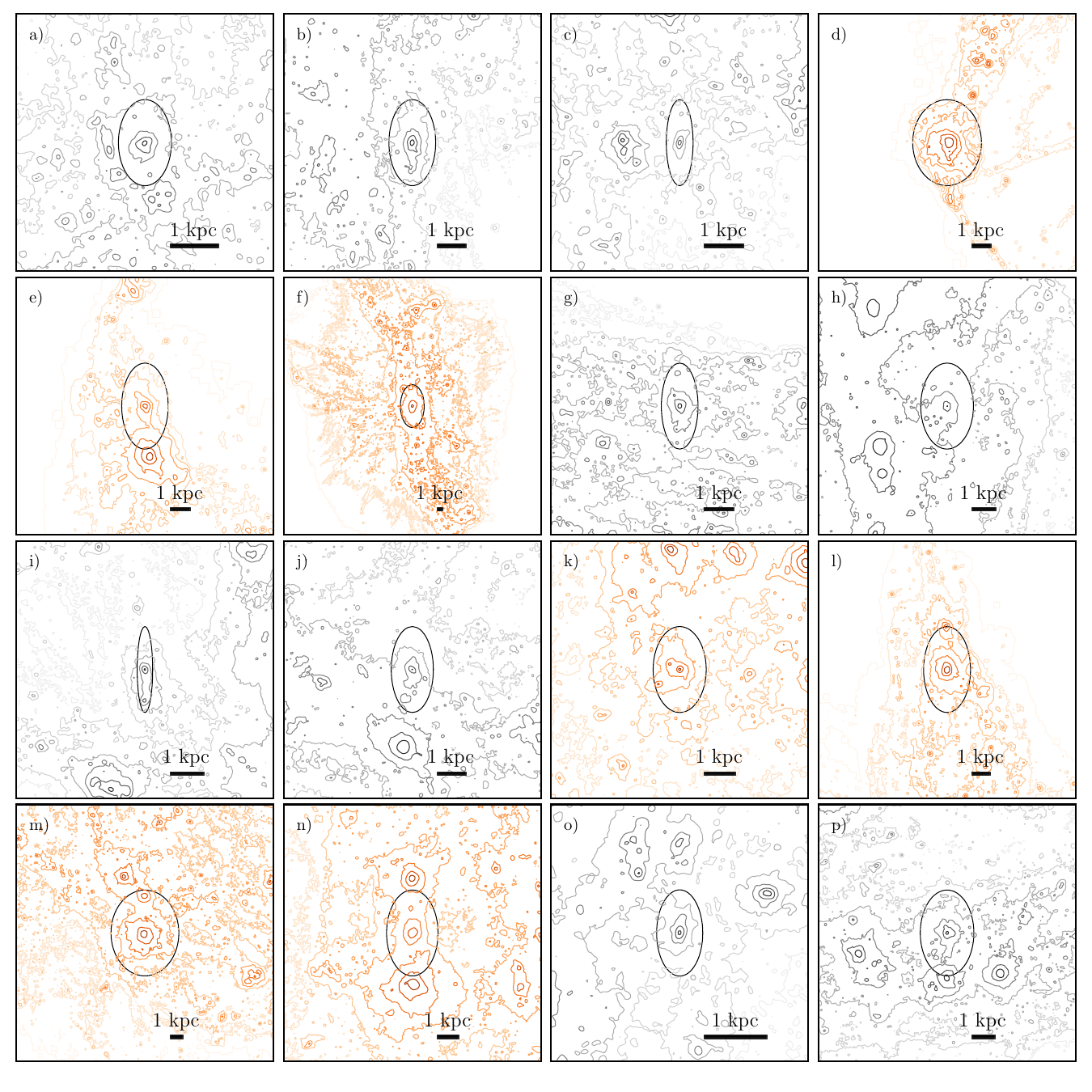}
    \caption{Series of dark-matter isodensity maps for a selection of 16 dark matter haloes. All haloes have been rotated to align the halo outer major and minor axes with the vertical and horizontal axes of the panel. The systems have then been projected along the intermediate axis. Orange iso-densities correspond to haloes that formed stars at $z=6.14$, while black to haloes without stars. In each panel we have superimposed an ellipse with semi-major and semi-minor axes $\rvir$ and $q\rvir$ long, respectively, with $q$ the outer minor-to-major axis. Panel f) shows the central halo.}
    \label{fig:alignedhalos}
\end{figure*}

Phenomena related to the regeneration of the cusp in haloes have been discussed, for instance, by \cite{Laporte2015} and \cite{Orkney2021}. \cite{Laporte2015} studied the merger trees for 4 out of 10 massive satellites in a MW-like dark-matter halo in the Aquarius Aq A-2 suite of simulation, and they found that a large fraction of haloes that undergo 1:3 down to 1:30 mergers are the most likely to reform a cusp at redshift $z=0$, due to the accretion of dense dark-matter structures able to reach the halo center before being tidally disrupted. In our case, the central cusps in the progenitor haloes are first erased by the ignition of star formation, which in the simulation starts at $z=16$ (see \citealt{Calura2022}). From $z=10$ to $z=8$, as the halo accretes mass, the core radius reduces, with a major effect during the mass accretion subsequent to minor mergers. The major merger at redshift $z\gtrapprox8$ produces, in the end, a halo with core size much smaller the ones of the progenitor haloes (see Fig.~\ref{fig:centralqst}). \cite{Orkney2021} also argues that mergers can provide a viable mechanism to heat and erase dark matter cusps. Although in their case the effect is stronger and effective in ultrafaint dwarf galaxies with quenched star formation, we do not find any sign of density core formation induced by mergers, since almost all haloes that did not form stars and all haloes in the DMO simulation at $z=6.14$ have dark matter cusps (see Fig.~\ref{fig:logscrcMvir}). 

The top row of panels in Fig.~\ref{fig:centralqst} shows the shape parameter profiles ($q$, $s$, and $t$) of the corresponding haloes in the bottom panels. Here, the profiles highlight the change of inner shape resulting from the growth of the stellar component. At early times ($z=10$ and $z=8$), the progenitor haloes are approximately prolate in the center, similarly to what measured in the DMO simulation or in the other starless haloes, with an inner triaxility $t\simeq0.65-0.8$ ($q\simeq0.25-0.5$ and $s\simeq0.7$). The subsequent mass accretion and the later episodes of star formation as the one at $t=0.7-0.8\Gyr$ (see Fig.~\ref{fig:SFH}) produce a significant redistribution of dark matter in the inner region resulting in an oblate/spherical distribution ($t\simeq0.25$, $q\simeq0.7$, $s\simeq0.9$).


\subsection{Prolateness of the outer haloes}
\label{subsec:prolate}

While baryons certainly impact the inner regions of dark matter haloes, the overall shape of both classes of haloes (those with and without stars) remains largely consistent in the outer regions, exhibiting a distinct prolate tendency towards $\rvir$. Here, the median outer triaxiality reads $\tmean=0.73_{-0.20}^{+0.24}$ in haloes with stars,  comparable to the $\tmean=0.81_{-0.25}^{+0.13}$ of the smaller dark companions (see Fig.~\ref{fig:shape}), with the exception that the latter are prolate at all scales, with no statistical change of triaxiality with radius (Fig.~\ref{fig:shapeprof}).
Most importantly, this same outer degree of prolateness is measured also in all haloes identified in the DMO simulation, which is a strong indication that whatever mechanism determines the larger scale structure of haloes, it is likely independent of baryonic matter (see right panel of Fig.~\ref{fig:shapeprof}).

\cite{Allgood2006} conducted a comprehensive study of the relationship between halo shape and mass and redshift of a wide range of haloes in six high-resolution dissipationless simulations in a $\Lambda$CDM Universe. They analyzed virial halo masses ranging from $9\times10^{11}-2\times10^{14} \h\Msun$ and a redshift range of 0–3 finding that most of the haloes tend to be prolate, with a relatively low average minor-to-major axis ratio ($\qmean\simeq0.6$) for the lowest mass bin and higher redshift, which is agreement with our results\footnote{Note that the estimates of shape of \cite{Allgood2006} are computed at $0.3\rvir$.}. In their study, the tendency to be prolate increases with the the halo mass, meaning that more massive haloes are more prolate in the outer regions. The prolateness of our dark-matter haloes is also in agreement with \cite{VeraCiro2011} who, using high-resolution cosmological $N$-body simulations from the Aquarius project \citep{Springel2008} find the shape of haloes evolves in time, from prolate shapes at high redshift to triaxial/oblate geometry at present day (see also \citealt{Despali2014}). In Fig.~\ref{fig:tvsMvir} we show the outer triaxiality parameter $t$ (top panel), the outer minor-to-major axis $q$ (middle panel) and $q_{0.75}$ (bottom panel), the minor-to-major axis computed within $[0.5,0.75]\rvir$ as measure of prolateness in an intermediate distance, as a function of the halo virial mass for all haloes classified within the full physics and the DMO simulations. Differently from \cite{Allgood2006}, we do not find any significant dependence of shape on halo mass at any distance, and neither find an increasing degree of prolateness (measured by $q$ for a comparison) for decreasing halo mass.

The prolate shape of dark-matter is supposed to have a strong correlation with the halo merger history. Prolate distributions reveal how mergers between haloes occur along preferred directions \citep{Faltenbacher2005,Zentner2005} or how the inflow of material along filaments shapes the outer halo. In Fig.~\ref{fig:alignedhalos} we show dark matter iso-density maps from a relatively large selection of haloes with and without stars (orange and black, respectively). In each panel, the systems have been rotated to align the halo's major and minor axes with the vertical and horizontal axes of the panel. We recall that, when computing halo shapes from dark matter particles, rather than using only particles classified as group members by \textsc{hdbscan}, we include in the analysis all particles enclosed within the circularised halo virial radius. This makes our estimates of halo shapes sensitive to the granularity of the background that is given, for instance, by sub-haloes in the process of merging.  As it can be seen from Fig.~\ref{fig:alignedhalos}, in most cases the selected haloes present a relatively large number of interacting sub-haloes distributed, preferentially, along the major axis (e.g. panels a, i, l, m, n). 

{As pointed out}, the least massive haloes in the simulations are, on average, dynamically younger than more massive ones, they are in their formation phase and, thus, more strongly affected by the frequent infalling of material from mergers. On the other hand, more massive haloes have had more time to relax and start to dominate the surrounding mass distribution, thus their elongation is hardly influenced by mergers with small mass ratios, but rather by accretion along filaments of the cosmic web, which provide a slow and continuous fueling along preferential directions. In this latter case, the ability of filaments of determining prolate outer shapes of haloes is expected to be much more important at high redshift, when the filament cross section is small compared to the size the halo \citep{VeraCiro2011}. In our case the effect is particularly evident looking at the central, most massive halo (panel f of Fig.~\ref{fig:alignedhalos}), clearly elongated towards the direction of a filament that flows mass in it, as it can be also appreciated from the dark-matter streaming velocity field shown in the bottom right panel of Fig.~\ref{fig:simz}).




\subsection{Discussion}
\label{subsec:lim}

Although the simulations analyzed in this work are innovative in many respects (e.g. implementation of individual star-formation, feedback from individual stars, sub-parsec resolution), processes such as radiative feedback from massive stars, molecular gas physics, or population III stars, have not been accounted for in the thermal and feedback models adopted. Therefore, in this Section, we briefly discuss whether and how our results could be modified by these processes, differentiating between: i) halo shapes; ii) core/cusp transformation.

\begin{itemize}
    \item[i)] We argued that the outer halo shape is mainly determined by the environment, specifically by mergers with smaller haloes and accretion of matter along filaments, as we showed comparing full physics and DMO. This result proved to be consistent with the findings of other authors who conducted similar analyses, especially those focusing on high-redshift halo shapes (see Section~\ref{subsec:prolate}). Thus, since the dark matter distribution on a large scale is predominantly insensitive to the inclusion of baryons, here we do not expect different implementations of the thermal model and/or the incorporation of radiative feedback, Pop III stars, or any other baryonic physics to significantly change our results.

    Regarding the inner halo shapes, we showed, instead, that the primary factor responsible for the sphericization is the condensation of baryons at the halo's center. The core of prolate/triaxial (dark matter) mass distributions are box orbits that are centrophilic and, thus, characterized by close passages toward the system's center \citep[see, e.g. ][]{Gerhard1985,Valluri1998,Merritt1999}. When slowly forming a compact massive object at the center of a galaxy, as in our case through the isotropic accumulation of baryons, \cite{Debattista2008} showed that the consequent steepening of the potential wells induces a redistribution of orbital families. The dark matter particles populating these orbits, which are more likely to reach the center, are scattered onto deformed box orbits, tube or loop orbits that have a rounder structure. As a result, the mass distribution also becomes rounder.


    This sphericization has been reported in a large variety of simulations, both cosmological and of isolated galaxies. For instance, studying isolated prolate/triaxial haloes, \cite{Valluri2010} confirmed that changes in halo shapes are driven by a regular adiabatic deformation of orbital families, with orbits that become rounder together with the global potential, but not chaotic \citep{Kalapotharakos2004,Shen2004}. \cite{Kazantzidis2010} delved deeper into investigating how the growth of a central disk galaxy impacts a triaxial dark matter halo, discovering that the maximum sphericization occurs when the symmetry axis of the disk aligns with the major axis of the halo. The thermal model of the cosmological Illustris galaxies of \cite{Chua2019} accounts for gas self-shielding and radiative feedback from massive stars. \cite{Chua2022} investigated different feedback models, modifying the AGB wind prescriptions and BH accretion modes. The NIHAO simulations \citep{Wang2015}, in part analyzed in \cite{Butsky2016}, are complemented by radiative feedback and a different implementation of delayed cooling. The EAGLE and Fenix simulations by \cite{Cataldi2021} include AGN feedback and metal-dependent radiative cooling, star formation, chemical and energetic supernovae feedback \citep{Springel2005b}. Despite different implementations and/or the inclusion of additional physical properties, all studies on inner halo shapes based on these simulations find consistent results: rounder haloes in the central parts.

    Thus, haloes become rounder at their center as they acquire baryonic mass, almost independently of the exact conditions under which they do so. Based on these studies, we do not believe sphericization would be impeded or overly modified by physical processes that are missing  in our simulations unless turning off cooling. The only modification to the thermal model we believe can have an effect is the inclusion of primordial molecular cooling, which (in the limit of the molecular dissociation by UV background) would probably accelerate the process, leading to rounder haloes above the molecular cooling halo threshold (rather than the atomic cooling threshold) and, thus, making haloes rounder earlier in time.




    \item[ii)] The effect of baryonic physics in the context of the core/cusp problem is highly controversial and subject to ongoing debate. In a $\Lambda$CDM paradigm, it is reasonably well established that alterations of the inner dark-matter density of haloes must be a result of some heating mechanism that gives energy to dark-matter particles, wiping out the central cusp.  Much, however, remains poorly understood, especially at the regime of the dwarf galaxies.

    The cosmological, $\Lambda$CDM, high resolution simulations of \cite{Governato2012}, differently from ours, also include H$_2$ creation, cooling and dissociation by Lyman-Werner radiation. They focus on $z=0$ galaxies, and most of the systems in their sample with stellar mass $<10^9\,M_{\odot}$ formed an extended core of dark matter. Similar results are obtained by \cite{DiCintio2014b} studying galaxies from the MUGS project \citep{Stinson2010} - a sample of 16 zoomed-in regions of cosmological simulations of L$^*$ galaxies. They explored variations of the initial mass function, the density threshold for star formation, and energy from supernovae and massive stars, and found that the inner slope of the dark matter density profile depends on the stellar-to-halo mass ratio, almost independently of the specific parameters adopted in their stellar feedback model, with cores forming also for moderate stellar-to-halo mass ratios (see also \citealt{Governato2012}). Note, however, that these works follow the entire evolution of the simulated galaxies and focus on the $z=0$ outcome. We are instead limited to $z=6.14$ and low-mass galaxies, making less trivial this comparison. Not to mention the very different resolution (which plays a very import role) and, most importantly, the fact that we model formation and feedback from individual stars. 
    
    Although in a non-cosmological context, but via simulations of isolated dwarf galaxies, \cite{Read2016} were able to reach resolutions similar to ours and to resolve individual supernovae explosions. They modeled a stellar age and mass dependent injection of energy, momentum, mass and heavy elements over time via SN II and Ia explosions. Most importantly, they included stellar winds and radiation pressure from massive stars. Gas is also set to a metallicity $10^{-3}$ solar, to mimic Pop III enrichment. They found that the least massive galaxies of their sample (virial mass $10^8M_{\odot}$) were able to build up a fully formed core in less than few Gyr, given a star formation activity sufficiently prolonged in time. On the contrary, \cite{Oman2015} could not find substantial signs of core formation in low mass galaxies. They studied systems from the high-resolution cosmological simulations from the EAGLE project (\citealt{Crain2015}; plus DMO analogous) which include star formation, stellar mass-loss, energy feedback from star formation, gas accretion on to and mergers of BHs, and AGN feedback. 

    Based on these examples (but see also \citealt{Boldrini2021b} for a review), we may expect, for instance, the addition of molecular gas cooling, or to the inclusion of radiative feedback to have a negligible effect. However, it is still hard to fully predict in which way modifications of the thermal and feedback models could really change our results. This is not just due the inclusion of addition processes, but also because the implementation of the ones already in place, as well as the resolution of our simulations, are, to some extent, unprecedented. As an example, the sub-parsec resolution pushes the star formation threshold up to $n\simeq10^5$ cm$^{-3}$, typical of giant molecular clouds, rather than $10^3$cm$^{-3}$ of typical simulations with lower resolution (see \citealt{Calura2022} for a discussion). As shown by \cite{Benitez2019}, core formation and the extent of the core is quite dependent on the star formation density threshold, with larger cores forming for larger star formation thresholds.
    Anyhow, we are currently incorporating additional physical properties into our sub-grid model, and we will conduct a detailed study to explore their effects in a forthcoming paper (Calura et al., in prep.).

\end{itemize}

\section{SUMMARY AND CONCLUSIONS}
\label{sec:conc}

In this paper we study how the structure of young dark matter haloes is influenced by their baryonic content and by environment during the initial stages of cosmic structure formation, when the Universe is $0.92\Gyr$ old. We analyze the output of a cosmological, zoom-in simulation aimed at reproducing the properties of a star forming complex observed at $z=6.14$, and an analogous DMO simulation in the context of the SIEGE project \citep{Calura2022}. We identify haloes within the simulations using the density-based, hierarchical clustering method \textsc{hdbscan}, we derive their density distributions and shapes diagonalizing the haloes' unweighted shape tensor with an iterative algorithm, and we fit the resulting ellipsoidal density profile with a flexible coreNFW model that allows to account for the possibility that the halo has a core of constant density in its central parts.

The very high resolution of the simulation enables us to cover the very low mass regime of the SHMR, in the mass range $\log\Mvir/\Msun=10^{7.5-9.5}$ which is in agreement with the very few relations found in the literature at the same redshift and, at most, covering the high mass end of our relation. In our specific case, we have quantified that, even if individual halo masses can change within a factor 1.6 when accounting for their triaxial shape, the effect on the overall SHMR is negligible.


Almost all haloes that have formed stars at $z=6.14$ have their cusp flattened into a core of constant density. We find that the mass threshold for the formation of the core is the same as the star formation threshold, about $\log\Mvir/\Msun\simeq7.6$, which corresponds to the mass of an atomic cooling halo at this redshift. In the DMO simulation we do not find signs of density core formation. This is a clear indication that, in our case, the dark matter cusp is heated by phenomena related to the injection of energy through baryonic processes, such as stellar feedback from winds and supernovae that become effective as soon as star formation begins. Additionally, we find that the extent of the core in the dark matter density is proportional to the mass and size of the formed stellar system. The only exception is the most massive halo in the simulation ($\log\Mvir/\Msun\simeq9.6$) which, instead, has one of the smallest cores. In agreement with previous works, after analysing the merger history of the halo, we attribute the attenuation to the phenomenon of cusp regeneration caused by mergers, indicating an intricate variety of processes that can concur in shaping the inner density distributions of haloes.

Also in this very low mass range, we confirm that baryonic infall of gas and baryonic feedback affects the distribution of dark matter at the centers of haloes making, on average, the inner shapes of haloes that have formed stars rounder than the inner shapes of systems that have not formed stars yet, and also rounder than corresponding haloes in the DMO simulation. For haloes that have formed stars in the simulation, which are also the most massive, we measure inner median intermediate-to-major and minor-to-major axes $\smean=0.84_{-0.12}^{+0.07}$ and $\qmean=0.66_{-0.13}^{+0.11}$, respectively, larger by 0.2 dex than the same quantities measured in the central parts of starless haloes ($\smean=0.67_{-0.19}^{+0.17}$ and $\qmean=0.48_{-0.11}^{+0.15}$) or haloes in the DMO simulation, extending results from previous studies to high redshifts and very low halo masses. We have quantified the degree of triaxiality of haloes via the one-dimensional triaxiality parameter and we found that all haloes (i.e. with and without stars in the full physics and in the DMO simulations) have the same degree of prolateness in the outer parts, to an extent that, at least for the mass range considered, does not depend on the halo mass. We have shown that the outer halo shape is predominantly determined by the environment: the elongated, prolate shape is driven by mergers with massive sub-haloes for the least massive haloes, and by the accretion of mass that occurs, preferentially, along dark matter filaments of the cosmic web for the most massive haloes in the simulations.

\section{ACKNOWLEDGMENTS}
We acknowledge support from PRIN INAF 1.05.01.85.01. AL acknowledges funding from MIUR under the grant PRIN 2017\-MB8AEZ.
This paper is supported by the  Fondazione ICSC, Spoke 3 Astrophysics and Cosmos Observations. National Recovery and Resilience Plan (Piano Nazionale di Ripresa e Resilienza, PNRR) Project ID CN\_00000013 "Italian Research Center on High-Performance Computing, Big Data and Quantum Computing" funded by MIUR Missione 4 Componente 2 Investimento 1.4: Potenziamento strutture di ricerca e creazione di "campioni nazionali di R\&S (M4C2-19)" - Next Generation EU (NGEU). This research was supported in part by Lilly Endowment, Inc., through its support for the Indiana University Pervasive Technology Institute. We acknowledge PRACE for awarding us access to Discoverer at Sofia Tech Park, Bulgaria. We thank the anonymous referee for useful suggestions that helped improving the quality of this work.

\section{DATA AVAILABILITY}
All data from the analysis of the simulation used in this article will be shared on request to the corresponding author.

\bibliography{paper}
\bibliographystyle{mnras}

\end{document}